\def\Nc{N_{\rm c}}
\def\q{{\bm q}}
\def\x{{\bm x}}

\def\tr{\operatorname{tr}}

\def\Im{\operatorname{Im}}
\def\grad{{\bm\nabla}}

\documentclass[prd,preprint,eqsecnum,nofootinbib,amsmath,amssymb,
               tightenlines,dvips]{revtex4}
\usepackage{graphicx}
\usepackage{bm}
\usepackage{amsfonts}

\def\zero{\mathsf{0}}
\def\one{\mathsf{1}}
\def\two{\mathsf{2}}
\def\three{\mathsf{3}}
\def\five{{\mathsf{5}}}

\def\zh{z_{\rm h}}
\def\Rads{\mathtt{R}}
\def\bul{\bullet}
\def\coeff{\varepsilon}

\begin {document}



\title
    {
      Coupling dependence of jet quenching in hot strongly-coupled
      gauge theories
    }

\author{
  Peter Arnold, Phillip Szepietowski, and Diana Vaman
}
\affiliation
    {%
    Department of Physics,
    University of Virginia, Box 400714,
    Charlottesville, Virginia 22904, USA
    }%

\date {\today}

\begin {abstract}%
{%
  Previous top-down
  studies of jet stopping in strongly-coupled QCD-like plasmas
  with gravity duals have been in the infinite
  't Hooft coupling limit $\lambda \to \infty$.  They have found
  that, though a wide range of jet stopping distances are possible
  depending on initial conditions,
  the maximum jet stopping distance $\ell_{\rm max}$ scales
  with energy as $E^{1/3}$ at large energy.  But it has always been
  unclear whether the large-coupling and high-energy limits commute.
  In this paper, we use the string $\alpha'$ expansion
  in AdS-CFT to study the corrections to the $\lambda{=}\infty$
  result in powers of $1/\lambda$.
  For the particular type of ``jets'' that we study,
  we find that (i) the naive expansion in $1/\lambda$ breaks down for
  certain initial conditions but (ii) the relative corrections to the
  {\it maximum}\/ stopping distance are small when $1/\lambda$
  is small.  More specifically, we find that the expansion in
  $1/\lambda$ is well behaved for jets whose stopping distance
  $\ell_{\rm stop}$ is in the range
  $\lambda^{-1/6} \ell_{\rm max} \ll \ell_{\rm stop} \lesssim \ell_{\rm max}$,
  but the expansion breaks down (and the fate of $\lambda{=}\infty$
  results is uncertain) for jets created in such a way that
  $\ell_{\rm stop} \ll \lambda^{-1/6} \ell_{\rm max}$.
  The analysis requires assessing the effects of all
  higher-derivative corrections to the supergravity action for the
  gravity dual.
}%
\end {abstract}

\maketitle
\thispagestyle {empty}


\section {Introduction and Results}
\label{sec:intro}

An important theoretical question for the study of jet quenching in
quark-gluon plasmas is how far a high-momentum low-mass excitation (such as
a high-energy massless parton with $E \gg T$) can propagate through the
plasma before losing its energy to the plasma and thermalizing.
A variety of authors
\cite{GubserGluon,HIM,CheslerQuark,adsjet,adsjet2,CHR}%
\footnote{
   See also Sin and Zahed \cite{SinZahed} for the earliest attempt we
   are aware of to discuss jet stopping in the context of gauge-gravity
   duality.
   See also ref.\ \cite{ZahedRecent}.
   In our paper, we will only consider analogs of light-particle jets
   and will not study the heavy-particle case.  For $\lambda{=}\infty$
   analysis of the latter, see, for example,
   refs.\ \cite{heavy1,heavy2} and references therein.
}
have used gauge-gravity duality to study this
problem in the strong coupling limit
$\lambda \equiv \Nc g_{\rm YM}^2 \to \infty$ of large-$\Nc$, ${\cal N}{=}4$
supersymmetric Yang-Mills (SYM) and related QCD-like plasmas.
The exact stopping distance depends on the exact
nature of the excitation, but it has been found that the maximum
stopping distance scales with energy as $E^{1/3}$ (in contrast to
the weak-coupling result, which is $E^{1/2}$ up to logarithms%
\footnote{
  A specific calculation for non-supersymmetric QCD
  of the stopping distance at weak coupling in
  the high-energy limit may be found in ref.\ \cite{stop}.
  However, the scaling of this result was
  implicit in the early pioneering work of
  refs.\ \cite{BDMPS,Zakharov} on
  bremsstrahlung and energy loss rates in QCD plasmas.
  The introduction of supersymmetry will not change the
  conclusion that the stopping distance scales as $E^{1/2}$ (up to
  logarithms) at weak coupling.
}%
).
It is an interesting and instructive result that the power of
energy depends on coupling,
but there is a potential loophole to this conclusion, which has to do
with orders of limits.  The work on this problem to date has implicitly
taken the $\lambda \to \infty$ and $\Nc \to \infty$ limits {\it first}, and
only then considered the limit $E/T \to \infty$ of high-energy
excitations.
Consider two possibilities for large but finite $\lambda$:
(i) the maximum stopping distance grows like $E^{1/3}$ for arbitrarily
large energies, versus
(ii) it first grows like $E^{1/3}$ for
$T \ll E \ll \lambda^n T$ (for some $n>0$) but
then behaves differently for $E \gg \lambda^n T$.
We cannot distinguish between these possibilities with only
$\lambda{=}\infty$ results.
Yet the distinction is an important one if there is any qualitative
application of such results to real QCD plasmas,
since $\lambda$ in the real world is not a fantastically large number.
Similar issues arises with $\Nc$: Might jet stopping be qualitatively
different for $E \gg \Nc^n T$?

In this paper, we will keep%
\footnote{
   For a discussion of one potential source of $1/\Nc$ corrections to jet
   propagation, see Shuryak, Yee, and Zahed \cite{SYZ}.
}
$\Nc{=}\infty$
and address the question of
what happens when $\lambda$ is large but not infinite.
In the gravity
dual, this will require considering the effect of
string corrections to the supergravity action.
These corrections correspond to higher-derivative terms
in the supergravity action,
such as the 4th power of the Riemann curvature.
Formally, the effects of higher and higher derivative corrections
to the supergravity action
are suppressed by more and more factors of
$1/\sqrt{\lambda}$, but these suppressions might be compensated by
large factors of $E/T$ in the jet stopping problem.

\begin {figure}
\begin {center}
  \includegraphics[scale=0.8]{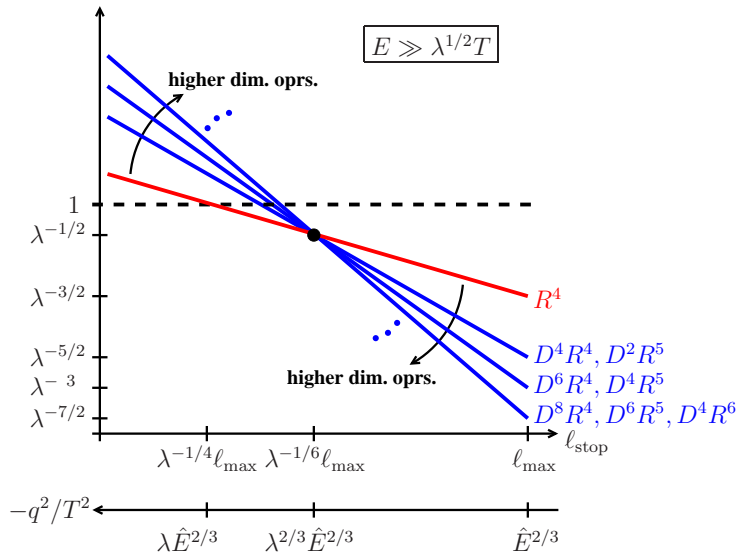}
  \caption{
     \label{fig:corrections}
     A parametric picture of the relative importance of
     higher-derivative corrections to the low-energy
     supergravity action as a function of the stopping distance
     $\ell_{\rm stop}$
     (using the $\lambda{=}\infty$ result for $\ell_{\rm stop}$).
     The axis are both logarithmic, and an importance of $1$ indicates
     that the individual correction would, by itself, significantly modify
     the $\lambda{=}\infty$ analysis.  Our measure of ``importance'' is
     explained in section \ref{sec:importance}.  Also shown, as an alternative
     horizontal axis, is the 4-dimensional virtuality $-q^2$ of
     the source that created the jet, where ${\hat E}\equiv E/T$.
  }
\end {center}
\end {figure}

For the particular type of ``jet'' excitations that we will study,
fig.\ \ref{fig:corrections} summarizes our results.
This figure depicts the parametric importance of corrections as a function
of the $\lambda{=}\infty$ result $\ell_{\rm stop}$ for the stopping
length of the jet.  Parametrically, the maximum scale for $\ell_{\rm stop}$
is
\begin {equation}
   \ell_{\rm max} \sim \frac{E^{1/3}}{T^{4/3}} \,.
\label {eq:lmax}
\end {equation}
The straight lines on this log-log plot represent simple power-law
dependencies on the stopping distance.
The first correction to the 10-dimensional low-energy
supergravity action for the gravity dual
to ${\cal N}{=}4$ super-Yang-Mills is of the form $R^4$
\cite{GrossWitten}
(plus other terms related by supersymmetry), where $R^4$ is
short-hand for a particular combination of contractions of four powers
of the Riemann tensor.  We've labeled each curve in
fig.\ \ref{fig:corrections} with a few examples of the type of
higher-derivative correction that contributes to each.
(As we will see, the importance to jet stopping of a correction term in the
supergravity action is determined by more
than just its
dimension, and we've only shown examples from among the most important
terms of each dimension.)
We will explain later exactly what we mean by the importance of
an operator, denoted by the vertical axis.  It is not quite the same
thing as the relative change in stopping distance due to that operator,
but, when the ``importance'' is small, the effect on the
stopping distance will also be small.
Finally, we note that fig.\ \ref{fig:corrections} assumes the
high-energy limit
$E \gg \lambda^{1/2} T$.  If $E \ll \lambda^{1/2} T$
(which is equivalent to $\lambda^{-1/6} \ell_{\rm max} \ll 1/T$), then the
corrections to $\lambda{=}\infty$ jet stopping results remain small
from $\ell_{\rm stop} \sim \ell_{\rm max}$ all the way down to
$\ell_{\rm stop} \sim 1/T$, which is the smallest
jet stopping distance that we will consider.%
\footnote{
  At a technical level, we define where the jet stops
  \cite{adsjet,adsjet2}
  following Chesler et al.\ \cite{CheslerQuark,CJK}
  as the location where the jet's energy and
  momentum and charge first begin to evolve hydrodynamically.
  Since hydrodynamics is an effective theory only on distance scales
  $\gg 1/T$ at strong coupling, it does not make sense to apply this
  definition to stopping distances small compared to $1/T$.
}

\begin {figure}
\begin {center}
  \includegraphics[scale=0.5]{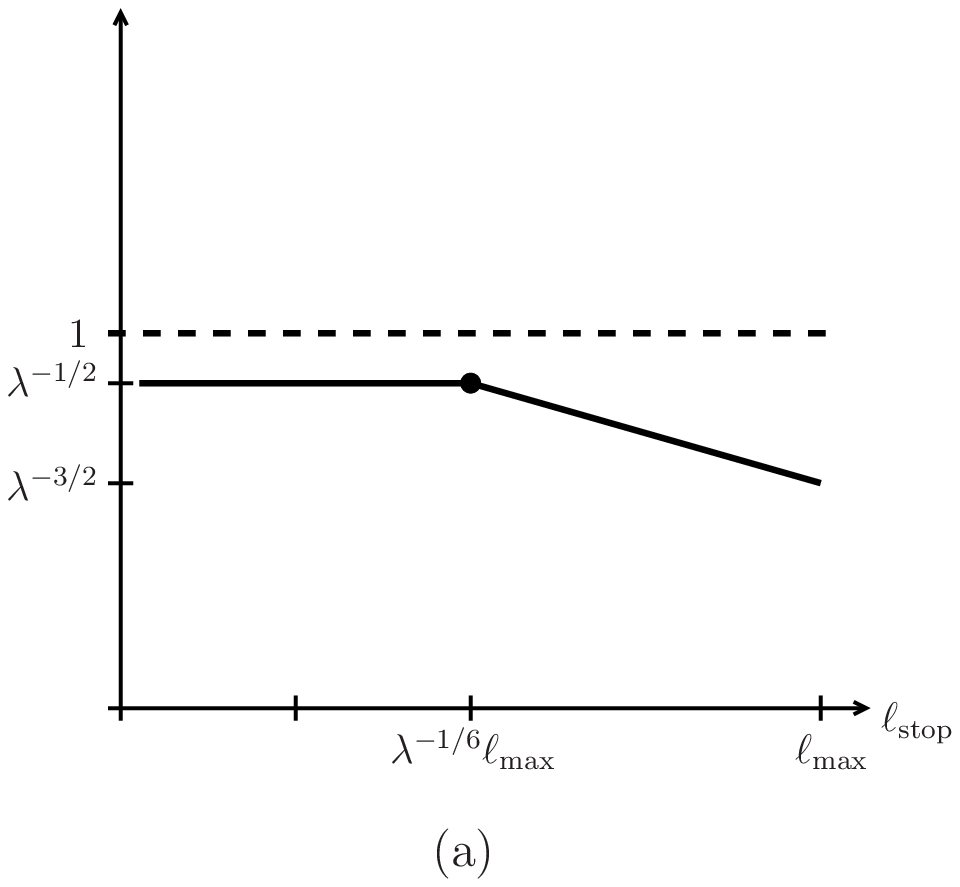}
  \includegraphics[scale=0.5]{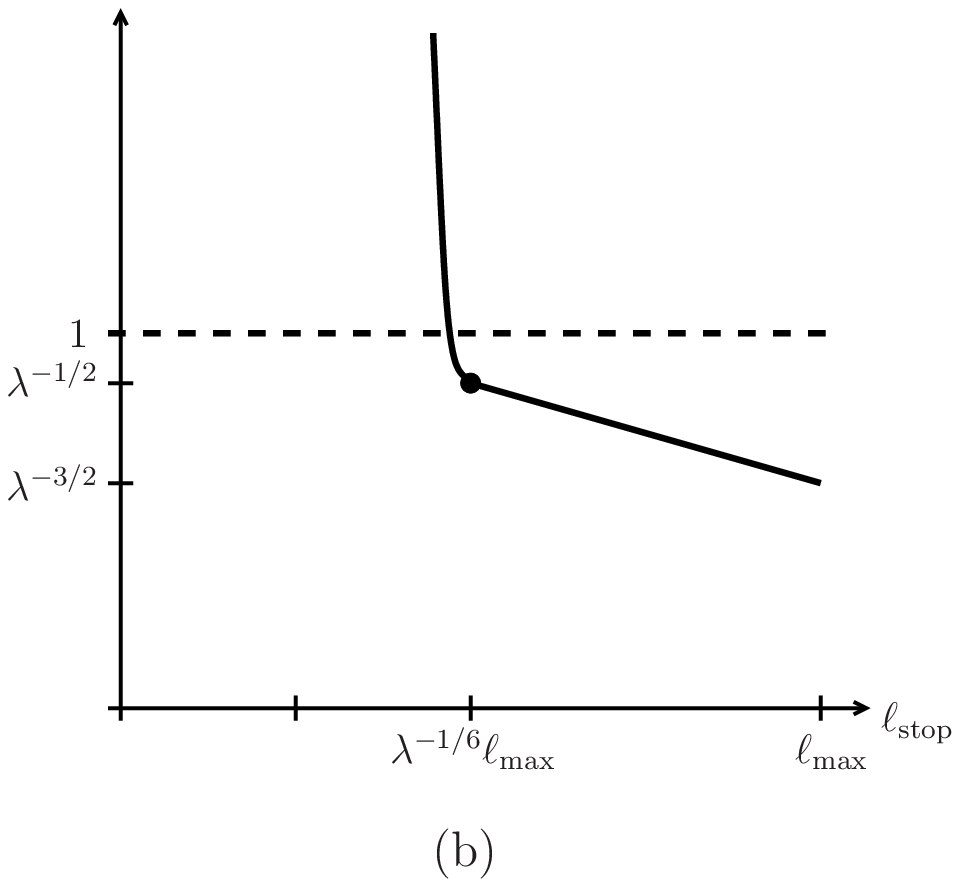}
  \includegraphics[scale=0.5]{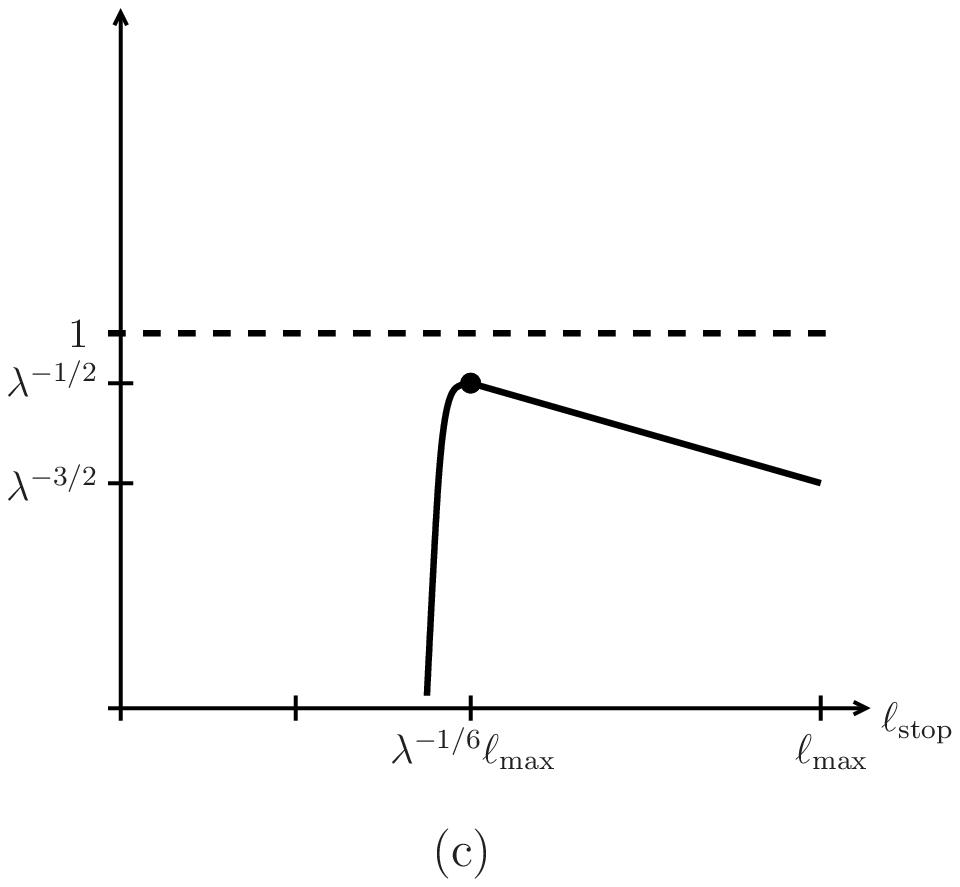}
  \caption{
     \label{fig:correctionABC}
     Like fig.\ \ref{fig:corrections} but showing some different
     behaviors that the total correction (summing all higher-derivative
     corrections) might conceivably have.
  }
\end {center}
\end {figure}

From fig.\ \ref{fig:corrections}, we conclude that corrections to the
usual $\lambda{=}\infty$ analysis of jet stopping are small for
$\ell_{\rm stop} \gg \lambda^{-1/6} \ell_{\rm max}$.
This includes
in particular the case of jets that travel the maximum stopping
distance,
$\ell_{\rm stop} \sim \ell_{\rm max}$, given by (\ref{eq:lmax}).
For jets with smaller stopping distances,
$\ell_{\rm stop} \ll \lambda^{-1/6} \ell_{\rm max}$,
the conclusion is uncertain.
As one decreases $\ell_{\rm stop}$ down from $\ell_{\rm max}$,
all the corrections become the same order at
$\ell_{\rm stop} \sim \lambda^{-1/6} \ell_{\rm max}$, but the
corrections are individually all ${\it small}$
(order $\lambda^{-1/2}$) at this scale.
At lower $\ell_{\rm stop}$, the expansion in higher-derivative
corrections is no longer useful, and one can imagine different
behaviors depending on how that expansion sums.
Perhaps the relative
size of the total correction to the $\lambda{=}\infty$ result
flattens out, as depicted in fig.\ \ref{fig:correctionABC}a.
Perhaps the corrections sum to give rapid (e.g.\ exponential)
growth, as in fig.\ \ref{fig:correctionABC}b.  Perhaps they sum
to give rapid suppression, as in fig.\ \ref{fig:correctionABC}c.
Figuring out what happens for
$\ell_{\rm stop} \ll \lambda^{-1/6} \ell_{\rm max}$
would presumably involve a full
string-theory analysis of the problem, which is beyond the scope
of this paper.  Here we just note that the validity of the
$\lambda{=}\infty$ results for
$\ell_{\rm stop} \ll \lambda^{-1/6} \ell_{\rm max}$ is unclear:
it might be that $\lambda{=}\infty$ remains a good approximation
in that case, as in figs.\ \ref{fig:correctionABC}a and c,
or fails completely, as in fig.\ \ref{fig:correctionABC}b.

We should emphasize at the outset that finding the {\it parametric}\/
dependence of corrections shown in fig.\ \ref{fig:corrections} will
not depend on knowing details of the precise form of higher-derivative
corrections to the supergravity action,
nor on details of their precise effects on
the AdS$_5$-Schwarzschild background.  Such details are not known for
corrections involving high powers of curvature.
And though we have taken care in fig.\ \ref{fig:corrections} to only depict
higher-derivative corrections that actually appear
as string corrections to Type IIB supergravity,%
\footnote{
  For a nice summary of higher-dimensional gravitational corrections
  in Type II supergravity generated by tree-level string amplitudes
  (i.e.\ in the $\Nc{=}\infty$ limit), see
  Table 1 of Stieberger \cite{Stieberger}.
  Though not relevant to the $\Nc{=}\infty$ case we are discussing,
  a nice discussion of corrections generated from one-loop string
  amplitudes may be found in Richards \cite{Richards}.
}
which is the case relevant to ${\cal N}{=}4$ SYM, our qualitative
results do not depend on these details either.
Our results will
follow from general arguments concerning
any expansion of the supergravity action in
higher-derivative corrections.

Our results in fig.\ \ref{fig:corrections} are parametric in nature.
In fact, given the types of ``jets'' that we will study, the
maximum stopping distance scale $\ell_{\rm max}$ given
by (\ref{eq:lmax}) will only be defined parametrically.
It is the distance scale beyond which the amount of charge that
a highly-penetrating
jet deposits in the medium, on average, begins to fall exponentially.
One may define a related scale $\ell_{\rm tail}$ by characterizing this
exponential fall-off as
\begin {equation}
   \mbox{deposition}(x^3)
   \sim
   \mbox{prefactor} \times e^{-x^\three/\ell_{\rm tail}}
   \qquad \mbox{for} \qquad
   x^\three \gg \ell_{\rm max}
\label {eq:falloff0}
\end {equation}
for a jet moving in the $x^\three$ direction.
Fig.\ \ref{fig:corrections} indicates that the expansion in
higher-derivative corrections should be well-behaved around
$\ell_{\rm max}$, and correspondingly the corrections to
$\ell_{\rm tail}$ should be well-behaved.  We will explicitly compute
the leading, $R^4$ correction to $\ell_{\rm tail}$.
The precise result depends on details of the type of source used to
initially create the jets.  As an example, here is the result
that we will find if we imagine (as a thought experiment
in the field theory) creating a jet in the ${\cal N}{=}4$ SYM
plasma via the decay of a high momentum, slightly off-shell
graviton:
\begin {equation}
   \ell_{\rm tail} =
      \ell_{\rm tail}^{\lambda{=}\infty}
      \bigl[1 + 47.162 \, \lambda^{-3/2} + O(\lambda^{-5/2})\bigr] .
\label {eq:hxytail0}
\end {equation}
This result for $\ell_{\rm tail}$ increases with decreasing $\lambda$.

The first part of this paper will be devoted to a general analysis
of all higher-derivative corrections, as shown in
fig.\ \ref{fig:corrections}.  For that analysis, it will be
easiest to focus on the case $\ell_{\rm stop} \ll \ell_{\rm max}$,
where it turns out that one may simplify the analysis by using
a geometric optics approximation.  (Parametric results for
$\ell_{\rm stop} \sim \ell_{\rm max}$ may then be obtained by
extrapolation.)
In section \ref{sec:basic}, we review our basic framework \cite{adsjet}
for creating high-momentum excitations (``jets'')
and the simplest version \cite{adsjet2} of the corresponding
calculation of $\lambda{=}\infty$ stopping distances.
(Similarities and differences with the methods of other authors
will be briefly summarized.)
We will then be in a position to explain the nature
of the ``importance'' measure sketched in fig.\ \ref{fig:corrections},
as well as some assumptions that we make in our analysis.
We also discuss the source we use in the field theory
problem to create our jets, which we choose in a way that simplifies
the discussion of the effects of higher-derivative corrections in the
gravity dual.
In section \ref{sec:R4}, we start by discussing the first
higher-derivative correction to the supergravity theory, which is an
$R^4$ term in the gravity action.
We show how this correction generates the corresponding (red) curve
in fig.\ \ref{fig:corrections}.
Section \ref{sec:D2nR4} then extends the analysis to higher
derivative terms of the form $D^{2n} R^4$ in the gravity action.
Section \ref{sec:Rm} moves on to higher powers of the
curvature, $D^{2n} R^m$.
In section \ref{sec:zbad}, we briefly present another
way of looking at when the expansion in supergravity corrections breaks down.

The second part of this paper
presents the calculation of the scale $\ell_{\rm tail}$
(\ref{eq:falloff0}) of exponential tails.
Section \ref{sec:tail} begins by explaining more carefully the
definition of $\ell_{\rm tail}$ and the situations that produce the
desired behavior (\ref{eq:falloff0}).
Precise extraction of the fall-off (\ref{eq:falloff0}) is beyond the
range of validity of the geometric optics approximation, but one
may instead extract results by solving for quasi-normal mode
solutions to the appropriate linearized supergravity equations of motion.
We carry out this calculation for the
leading correction to $\ell_{\rm tail}$ in powers of $1/\lambda$.
For this particular calculation, in order to obtain the
specific result (\ref{eq:hxytail0}), we generalize to a slightly wider
selection of jet sources than the ones used to simplify the
general analysis of yet-higher-order corrections and
fig.\ \ref{fig:corrections}.

Finally, we conclude in section \ref{sec:conclusion} by
pointing out a basic and still open problem about how
$\ell_{\rm max} \propto E^{1/3}$
at strong coupling interpolates to $\ell_{\rm max} \propto E^{1/2}$
at small coupling.




\section {Basic framework}
\label {sec:basic}

\begin {figure}
\begin {center}
  \includegraphics[scale=0.5]{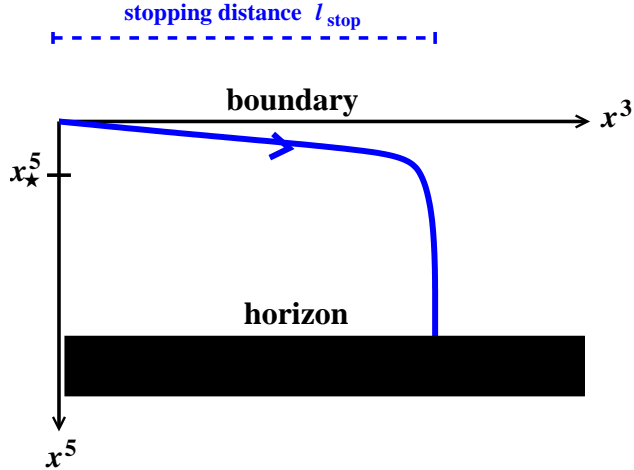}
  \caption{
     \label{fig:fall}
     Qualitative sketch of the motion of a particle (wave packet) through
     AdS$_5$-Schwarzschild.  As measured by $x^\zero$, the particle
     takes infinitely long to reach the horizon.
     Of special importance is the parametric
     scale $x^\five_\star$ in the fifth
     dimension, where the trajectory turns over and beyond which
     progress in
     $x^\three$ rapidly slows to a stop.
     Given our specific choice (\ref{eq:metric}) of coordinates,
     we often refer to $x^\five_\star$ as $z_\star$.
  }
\end {center}
\end {figure}

Our basic framework for studying jet stopping will be to create
high-momentum excitations of the strongly-coupled plasma by
perturbing the plasma with high-momentum sources, as in
refs.\ \cite{adsjet,adsjet2}, and studying the response.
As we will review below, for certain types of sources applied
to the quantum field theory, the response in the gravity dual is
the generation of a highly-localized and highly-oscillatory wave
packet that moves through space while falling in the fifth dimension
towards the black brane horizon.  This wave packet has approximately
well-defined 5-dimensional position and momentum, and (for $\lambda{=}\infty$),
its motion can be approximated (up to parametrically small corrections)
by a geodesic---that is, by the trajectory that
a particle would take through the AdS$_5$-Schwarzschild background.
This particle (geometric optics)
approximation makes the $\lambda{=}\infty$ calculation
of stopping distances particularly simple and efficient.
We will take the 3-momentum of our excitations to be in the
$x^\three$ direction, and
the stopping distance is simply given by how far in $x^\three$
the corresponding geodesic travels before falling into the black
brane horizon, as depicted in fig.\ \ref{fig:fall}.%
\footnote{
   For more discussion of why the distance the excitation travels before
   falling into the horizon should be identified with the stopping
   distance in the 3+1 dimensional field theory problem, see the
   discussion in ref.\ \cite{adsjet2}, as well as earlier discussions
   in the context of falling classical strings \cite{GubserGluon,CJK}
}

We will later study the effect of higher-derivative corrections to
the supergravity action by studying
their effects on trajectories such as fig.\ \ref{fig:fall}.
One effect is the change that higher-derivative corrections make to
the AdS$_5$-Schwarzschild background, but we will see that the dominant
effects are the changes they make to the equation of motion of the wave
packets, which will no longer follow geodesics.


\subsection {Notation}

In this paper, we will use Greek letters for 4-dimensional space-time indices
($\mu,\nu={\mathsf{0,1,2,3}}$).
Lower-case roman letters ($a,b$) will be used for 10-dimensional indices.
The first five of those 10 dimensions, corresponding to
AdS$_5$-Schwarzschild when $\lambda{=}\infty$, will be represented by
upper-case roman letters ($I,J={\mathsf{0,1,2,3,5}}$).  The remaining five
dimensions, corresponding to the compact 5-sphere $S^5$, will be
indicated by dotted lower-case roman letters ($\dot a, \dot b$).
When we use the adjective ``five-dimensional'' without further
qualification, then we are referring to the first 5
dimensions---those of AdS$_5$-Schwarzschild.

The form of the metric we will use for AdS$_5$-Schwarzschild is%
\footnote{
  The coordinate used in refs.\ \cite{adsjet,adsjet2}
  is $u = z^2/\zh^2$, which is $u = (z/2)^2$ when working in
  the units $2\pi T=1$ used there.
}
\begin {equation}
   (ds)^2 = \frac{\Rads^2}{z^2} \bigl[-f (dt)^2 + (d\x)^2 + f^{-1} (dz)^2\bigr]
   ,
\label {eq:metric}
\end {equation}
where $z$ is the coordinate $x^\five$ of the fifth dimension,
$\Rads$ is the radius of the 5-sphere (and will drop out of final results),
\begin {equation}
   f \equiv 1 - \frac{z^4}{\zh^4} \,,
\end {equation}
the boundary is at $z{=}0$, and the horizon is at
\begin {equation}
   \zh = \frac{1}{\pi T} \,.
\label {eq:zh}
\end {equation}
We will not need to worry about the details of regularizing the location
of the boundary in this work.%
\footnote{
   See section V of ref.\ \cite{adsjet2} for some discussion
   of boundary regularization in the
   context of making the particle approximation.
}

The metric of 4-dimensional flat space-time will be represented by
$\eta_{\mu\nu} \equiv \operatorname{diag}(-1,+1,+1,+1)$.


\subsection {Review of \boldmath$\lambda{=}\infty$ results}

\subsubsection {Set up}

In this paper, we follow the approach of
refs.\ \cite{adsjet,adsjet2} for studying the jet stopping
problem.  An external source is applied to
the strongly-interacting gauge theory in order to create the
initial high-energy high-momentum excitation.
Specifically, we add a source term to the Lagrangian,
\begin {equation}
   {\cal L} \to
   {\cal L} + {\cal N} \, O(x) \, e^{i\bar k\cdot x} \, \Lambda_L(x) ,
\label {eq:source}
\end {equation}
where ${\cal N}$ is an arbitrarily small source amplitude,
$O(x)$ is a source operator,
\begin {equation}
   \bar k^\mu \simeq (E,0,0,E)
\end {equation}
is the large 4-momentum of the desired excitation, and
$\Lambda_L(x)$ is a slowly varying envelope function that localizes
the source to within a distance $L$ of the origin
in both $x^\three$ and time.  For example,
\begin {equation}
   \Lambda_L(x) = e^{-\frac12 (x^\zero/L)^2} e^{-\frac12 (x^\three/L)^2}
   .
\label {eq:LambdaL}
\end {equation}
$L$ is chosen large compared to $1/E$ but small compared to the
stopping distance we wish to measure.
The small amplitude ${\cal N}$ is so that we can treat the external
source as a small-perturbation to the strongly-interacting gauge theory,
so that the source will never create more than one
jet with energy $E$ at a time.

The source operator $O(x)$ is a matter of choice.
As an example, ref.\ \cite{adsjet} found it convenient to focus on
``jets'' created by an external R-charge field (somewhat analogous
to the excitation that would be created by the hadronic decay of a
high-momentum W boson inside a standard-model quark-gluon plasma,
but with isospin replaced by R charge).
In that case the operator was $O(x) = j^\perp(x)$, where
$j^\mu$ is a combination of R current operators.
The details of the choice of source operator $O$ are unimportant
\cite{adsjet2}, however, as long as the operator has finite conformal dimension
in the $\lambda{=}\infty$ limit.  We'll later discuss the choice we
find convenient for the present study of finite-$\lambda$ corrections.

Sometimes in previous work \cite{adsjet}, the characteristic 4-momentum
$\bar k$ of the source has been taken to be exactly light-like, $\bar k
= (E,0,0,E)$.  Because the source is confined to a space-time
region of size $L$,
the momentum components $q^\mu$ of the source are smeared out around
$\bar k^\mu$ by an amount of order $1/L$, and so the typical magnitude
of the virtuality $q^2 \equiv q^\mu q_\mu$ of the source is then of order
$|q^2| \sim E/L \ll E^2$.

In later work, a calculational and conceptual
simplification was found if
one instead chooses the characteristic 4-momentum $\bar k$ to
be just a little bit time-like,
\begin {subequations}
\label{eq:koffshell}
\begin {equation}
   \bar k = (E+\epsilon,0,0,E-\epsilon)
\end {equation}
with
\begin {equation}
   \frac{1}{L} \ll \epsilon \ll E .
\label {eq:Lbound}
\end {equation}
\end {subequations}
The first inequality guarantees that the uncertainty in momentum does
not overwhelm the size of $\epsilon$.  In this case, the source
has an approximately well-defined virtuality in 4-momentum space
$q$ of
\begin {equation}
   -q^2 \equiv -q^\mu q_\mu \simeq -\bar k^\mu \bar k_\mu
   \simeq 4 \epsilon E.
\label {eq:virtuality}
\end {equation}
This is the case where the response
created in the dual gravity theory
can be shown \cite{adsjet2} to be a highly localized, highly
oscillatory wave packet that falls in the fifth dimension toward
the black brane horizon.
The trajectory of the wave packet is the geodesic that would be
followed by a massless 5-dimensional {\it particle}\/ traveling
in the AdS$_5$-Schwarzschild background as in
fig.\ \ref{fig:fall}.
Calculations using this particle picture \cite{adsjet2}
are much simpler and more efficient than calculations directly
in terms of the 5-dimensional field excitations \cite{adsjet}.


\subsubsection {The geodesic}

The 5-dimensional mass $m$ associated with a supergravity field, and therefore
with the 5-dimensional particle trajectory, is determined by the
conformal dimension $\Delta$ of the field theory operator dual to that field.%
\footnote{
  It is important to note that the masses of 5-dimensional fields
  in the gravity dual have nothing to do with the masses of
  4-dimensional excitations in the ${\cal N}{=}4$ SYM field
  theory.  The 5-dimensional mass $m$ is not the ``mass of a jet.''
}
We will take $\Delta$ to be of order one.
It was shown in ref.\ \cite{adsjet2} that, in the high-energy limit,
this mass does not affect the stopping distance for sources
described by (\ref{eq:koffshell}) when
$\ell_{\rm stop} \ll \ell_{\rm max}$, which will be our focus here.
So we may ignore the 5-dimensional
mass and focus on the trajectories $dx^I dx_I = 0$ corresponding to
null geodesics in AdS$_5$-Schwarzschild.  The solution for such
geodesics (for a metric
that depends only on $x^\five$ and has
4-dimensional parity) is%
\begin {equation}
   x^\mu(x^\five) = \int \sqrt{g_{\five\five}} \, dx^\five \>
      \frac{ g^{\mu\nu} q_\nu }
           { (-q_\alpha g^{\alpha\beta} q_\beta)^{1/2} }
   \,,
\label {eq:null}
\end {equation}
where the 4-momentum $q_\alpha$ with lower index is conserved
in 5-dimensional motion and is
given by the 4-momentum (\ref{eq:koffshell}) of our source,
$q_\alpha = \eta_{\alpha\beta} \bar k^{\beta}$.

Taking the integral in (\ref{eq:null}) all the way to the
horizon for $\mu{=}3$ (the direction of the jet), and
using the metric (\ref{eq:metric}),
gives the geometric optics approximation to the stopping distance
\begin {equation}
  \ell_{\rm stop} \simeq
  \int_0^{\zh} dz \> \frac{|\q|}{\sqrt{-q^2 + \frac{z^4}{\zh^4} |\q|^2}}
  ,
\label {eq:stop1}
\end {equation}
where we have used rotation invariance to rewrite $q_\three$ as $|\q|$.
Here and throughout this paper
we use the symbol $q^2$ for the 4-virtuality of
the source,
\begin {equation}
   q^2 \equiv q_\mu \eta^{\mu\nu} q_\nu < 0.
\end {equation}
Throughout this paper, we will restrict attention to the case
$-q^2 \ll E^2$ as in (\ref{eq:virtuality}), as this is the case which generates
stopping distances large compared to $1/T$.  In this limit,
the integral in (\ref{eq:stop1}) is dominated by small values of
$z$, of order
\begin {equation}
   z_\star
   \sim \zh \left( \frac{-q^2}{|\q|^2} \right)^{1/4}
   \sim \zh \left( \frac{-q^2}{E^2} \right)^{1/4}
   \ll \zh ,
\label {eq:zstar}
\end {equation}
corresponding to the parametric scale labeled $x^\five_\star$ in
fig.\ \ref{fig:fall}.
Neglecting parametrically small corrections, we may replace the
upper limit of integration by infinity in (\ref{eq:stop1}) to get%
\footnote{
   See ref.\ \cite{adsjet2} for this explicit result, but the
   parametric behavior $\ell_{\rm stop} \sim (E^2/{-}q^2)^{1/4}$,
   within its range of validity,
   was found earlier by Hatta, Iancu and Mueller
   \cite{HIM}.
}
\begin {equation}
  \ell_{\rm stop} \simeq
  \frac{\Gamma^2(\frac14)}{(4\pi)^{1/2}} \left( \frac{E^2}{-q^2} \right)^{1/4}
  \frac{1}{2\pi T}
  \,.
\label {eq:stop}
\end {equation}

The validity of (\ref{eq:stop})
is restricted to the range of validity of the geometric
optics approximation.  For a detailed discussion, see ref.\
\cite{adsjet2}.  Here, for the sake of simplicity of this review, we will
just give a quick, crude way to see the limit of applicability
from the result (\ref{eq:stop}) itself.
By the uncertainty principle, the components of the source's 4-momentum
will be smeared out by $1/L$, where $L$ is the source size.
Consequently, the virtuality $-q^2$
given by (\ref{eq:virtuality}) will only be (approximately) well-defined
when $\epsilon \gg 1/L$ and so when
\begin {equation}
   -q^2 \gg \frac{E}{L} \,.
\label {eq:q2lim}
\end {equation}
But the result (\ref{eq:stop}) from the geometric optics approximation
is only meaningful if $-q^2$ is approximately well defined.
Combining (\ref{eq:stop}) and (\ref{eq:q2lim}) requires
\begin {equation}
   \frac{\ell_{\rm stop}^4 T^4}{E} \ll L \,.
\end {equation}
On the other hand, it wouldn't be sensible to try to measure a
stopping distance unless we choose a source size that is smaller than
the distance we wish to measure.  So $L$ needs to satisfy
\begin {equation}
   \frac{\ell_{\rm stop}^4 T^4}{E} \ll L \ll \ell_{\rm stop} \,.
\end {equation}
Choosing such an $L$ is possible exactly when
$\ell_{\rm stop} \ll \ell_{\rm max}$ with $\ell_{\rm max}$ given
by (\ref{eq:lmax}).

The chance of propagating excitations created by sources like
(\ref{eq:source}) to distances $\gg \ell_{\rm max}$ is negligible.  But
showing this convincingly requires abandoning the geometric optics
analysis and doing a wave analysis, as in refs.\ \cite{HIM,adsjet}.
We leave that to section \ref{sec:tail}.
For most of this paper, we will stick to the region
$\ell_{\rm stop} \ll \ell_{\rm max}$ where the geometric optics
approximation is valid, and then make only parametric extrapolations
to the boundary $\ell_{\rm stop} \sim \ell_{\rm max}$ of the
range of validity.

We should also mention that even for $\ell_{\rm stop} \ll \ell_{\rm max}$,
the geometric approximation eventually breaks down at sufficiently large
$z \gg z_\star$.  At that point, however, the wave packet is falling
essentially straight down towards the horizon, and the fact that it can
no longer be treated as a particle no longer matters to how far it
travels in $x^\three$.
The stopping distance is determined by the behavior of the trajectory
for $z \sim z_\star$.  (That is, $z \ll z_\star$ and $z \gg z_\star$
give parametrically small contributions to the stopping length.)


\subsubsection {Other authors' methods for describing ``jets''}

Let's pause a moment to compare and contrast the type of jets we
create with some of the others studied in the literature.
Our formalism \cite{adsjet,adsjet2}
(with and without the particle approximation)
can be thought of as a concrete way to realize early
ideas of Hatta, Iancu and Mueller \cite{HIM}.

Alternatively,
there is a long history of considering jet-like states that are
dual to classical strings falling towards
the horizon in the gravity theory
(as well as a history of using
geodesics to help understand the strings' motion)
\cite{GubserGluon,HIM,CheslerQuark}.
One difference is that
the maximum stopping distance for these states is
parametrically smaller than for the states we consider ---
$\ell_{\rm max}$ for the states related to classical strings
scales as
$\lambda^{-1/6} E^{1/3} T^{-4/3}$ rather than the
$E^{1/3} T^{-4/3}$ of (\ref{eq:lmax}).  It's amusing to note
that, perhaps coincidentally, $\lambda^{-1/6} E^{1/3} T^{4/3}$ is the same
stopping distance scale where corrections become problematical in our
fig.\ \ref{fig:corrections}.  In any case, we will not attempt
here to study $1/\lambda$ corrections to
previous results based on classical strings.

Yet another, recent method for creating a gluon-like jet is to
generate it as a beam of synchrotron radiation from a heavy quark that is
forced into circular motion \cite{CHR}.
These gluon-like jets (under certain conditions)
penetrate a
distance of order $\ell_{\rm max}$ given by (\ref{eq:lmax}).
We will not attempt to study the $1/\lambda$ corrections in
this synchrotron problem, but we would not be surprised if they
work out similar to the $\ell_{\rm stop} \sim \ell_{\rm max}$ case in our
analysis.

Finally, since coupling does not run with scale in ${\cal N}{=}4$ SYM, we
are treating the coupling as large at {\it all}\/ scales relevant to energy
loss.  This is in contrast to programs, such as ref.\ \cite{LRW},
that try to isolate the soft effects of a strongly-coupled
medium on weakly-coupled hard bremsstrahlung or
pair-production vertices.%
\footnote{
  For a very brief summary of the relevant scales for the coupling, see, for
  example, ref.\ \cite{qm11}.
}
For work on $1/\lambda$ corrections in that context, see
ref.\ \cite{AEM}.


\subsection{Determining the importance of corrections}
\label {sec:importance}

We wish to check whether or not higher-derivative corrections
to the supergravity action can invalidate the $\lambda{=}\infty$ result for
finite but large $\lambda$ and large energy.
We have seen above that the $\lambda{=}\infty$
stopping distance is generated by the
behavior of the particle trajectory for $z \sim z_\star$
given by (\ref{eq:zstar}).
So the simple way to address our question is to check whether or not
higher-derivative corrections make significant changes to the
trajectory for $z \sim z_\star$.
The ``importance'' represented by the vertical axis
of fig.\ \ref{fig:corrections} is just the
relative effect on the trajectory at $z \sim z_\star$.

\begin {figure}
\begin {center}
  \includegraphics[scale=0.29]{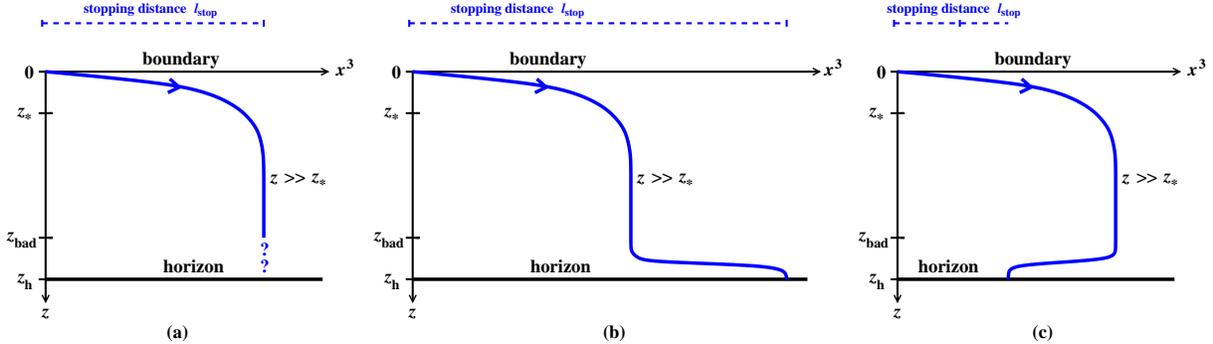}
  \caption{
     \label{fig:fall2}
     (a) A picture of the trajectory of the 5-dimensional wave packet
     when higher-derivative corrections have negligible effect on the
     trajectory at $z \sim z_\star$.  The trajectory stops moving
     in $x^\three$ as $z$ increases well above $z_\star$,
     falls straight toward the horizon, and then
     eventually reaches a large enough $z$ (labeled $z_{\rm bad}$)
     that the expansion in
     higher-derivative effects breaks down.
     (b,c) What we assume does not happen.
     In all of these figures, only the parametric ordering
     $z_\star \ll z_{\rm bad}$ is significant: $z_{\rm bad}$ is not
     necessarily close to the horizon.
  }
\end {center}
\end {figure}

We will see later in section \ref{sec:zbad}
that the relative effects of higher-derivative
corrections increase with increasing $z$.
We will see that, at sufficiently
high energies, there is always a point $z \gg z_\star$ where, in
the geometric optics approximation, the expansion
in effects of higher-derivative corrections goes bad.
In some cases, this will occur for $z$'s large enough that the
geometric optics approximation has already broken down there anyway.
But in all cases, we will make the following, physically reasonable
assumption in our analysis:
\begin {quote}
  {\it Assumption 1.} Once the 5-dimensional wave packet has
  stopped moving significantly in $x^\three$, so that it is falling essentially
  straight toward the horizon, then it will thereafter continue
  falling essentially straight toward the horizon and will not
  move significantly in $x^\three$ again.
\end {quote}
That is, if the wave packet's trajectory looks like fig.\
\ref{fig:fall2}a, we will assume that it does not actually behave like
figs.\ \ref{fig:fall2}b or c.

\begin {figure}
\begin {center}
  \includegraphics[scale=0.29]{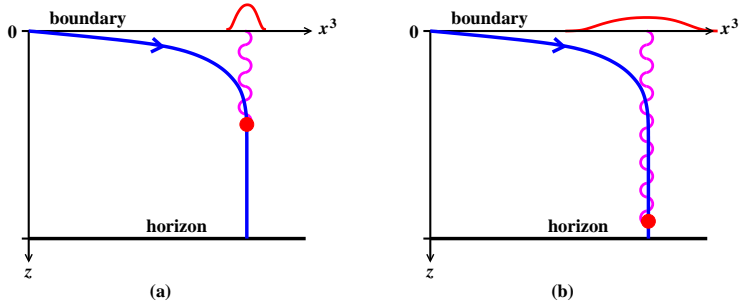}
  \caption{
     \label{fig:fall3}
     As the wave packet falls towards the horizon, the effects of its
     R charge on the boundary theory redshifts, corresponding to
     a hydrodynamically diffusing charge density measured in the
     4-dimensional plasma.
  }
\end {center}
\end {figure}

There is an additional assumption in our analysis.  We will
read off the jet stopping distance to be the distance $x^\three$ that
the 5-dimensional wave packet travels before it falls into the horizon.
But really, one should discuss the stopping distance directly in terms
of observables in the boundary theory.
Following Chesler et al.\ \cite{CJK},
we could look at the location of the source
terms for late-time hydrodynamic diffusion or sound propagation of
the charge and energy and momentum densities, deposited by the
jet in the medium once the jet stops and thermalizes.  For simplicity,
consider the case of jets that carry R charge.  The presence of
the wave packet in the bulk produces a response in the boundary fields
that are dual to R charge,
as depicted in fig.\ \ref{fig:fall3}a.  As the 5-dimensional wave packet
approaches the horizon, this disturbance is further and further
red-shifted, which corresponds to hydrodynamic diffusion of the
charge density in the boundary theory, as depicted at later times
in fig.\ \ref{fig:fall3}b.  In the $\lambda{=}\infty$ calculation, the
charge distribution is centered in $x^3$ over the position of the
wave packet as it approaches the horizon.  For the particular type
of jets we study in this paper, more detail may be found by comparing
refs.\ \cite{adsjet} and \cite{adsjet2},
but the same type of behavior occurs in
earlier works on jets dual to classical strings
\cite{GubserGluon,CheslerQuark,CJK}.
Our assumption will be that this correspondence continues
for high-energy jets at large but finite $\lambda$:
\begin {quote}
  {\it Assumption 2.} $1/\lambda$ corrections do not significantly
  modify the (approximate) equality between (i) the late-time $x^\three$
  position of the 5-dimensional wave packet as it approaches the horizon
  and (ii) the position where the jet stops and thermalizes in the
  4-dimensional field theory as measured,
  for example, by the center of the late-time diffusing distribution
  of R charge.
\end {quote}
In particular, in this paper we will not attempt to make a thorough
analysis of $1/\lambda$ corrections to the coupling between the
5-dimensional wave packet and the 5-dimensional gauge field
bulk-to-boundary propagator associated with making a late-time
measurement of the R charge distribution.%
\footnote{
  As discussed in ref.\ \cite{adsjet}, the bulk-to-boundary propagator
  associated with the late-time measurement is a low-momentum
  propagator.  This propagator therefore
  contains no powers of large energy $E$ that
  could compensate powers of $1/\lambda$, and so the corrections to that
  propagator are always small for large finite $\lambda$.
  When one extracts from the hydrodynamic response the place in the
  plasma where the charge was deposited, by applying the
  4-dimensional diffusion operator $(\partial_t - D_R \grad^2)$ as
  in refs.\ \cite{CJK,adsjet},
  the 5-dimensional bulk-to-boundary propagator just discussed
  gets truncated and only has support at the
  4-position of the bulk vertex.
  (See, for example, the related discussion in sec.\ III.D
  of ref.\ \cite{jj}.)
  Formally, that means that charge deposition only has
  support for 4-positions traveled through by the high-energy 5-dimensional
  wave packet created by the source.
  One might think this guarantees the validity of our Assumption 2.
  However, it is possible for a series of local functions to add
  up to something with support elsewhere, as in
  $\sum_{n} (a^n/n!) \, d^n\delta(x)/dx^n = \delta(x+a)$.
  Our Assumption 2 is that at late times (when the wave packet
  is very near the horizon),
  $1/\lambda$ corrections to the bulk 3-point
  vertex do not somehow sum up in a similar way to produce a large
  percentage change to the stopping distance.
}
(Nor will we make the
corresponding analysis for the coupling between the wave packet and
the graviton propagator associated with making a measurement
of the sound waves produced by the jet.)  We will
only analyze corrections to how far the 5-dimensional
wave packet travels.

\subsection{Our choice of source operator}
\label {sec:sourceO}

The source operator $O(x)$ used to generate the jet via
(\ref{eq:source}) determines the type of field that is excited in
the dual theory---that is, the type of wave packet whose trajectory
is depicted by fig.\ \ref{fig:fall}.
To keep our analysis simple, we will choose a class of source
operators that are dual to fields which are 5-dimensional scalars
after the $S^5$ reduction.
(We will slightly relax this restriction later, in section
\ref{sec:tail}, when presenting an example of an explicit calculation
of a leading correction.)
Moreover, in 10-dimensional language, it
will be convenient to mainly focus on the purely gravitational terms
in the supergravity action (about which the most is known).
So we will pick our 5-dimensional scalar field $\phi$ from the
components $h_{\dot a\dot b}$ of the 10-dimensional metric
fluctuations $h_{ab}$ with indices $\dot a$,$\dot b$
along the 5-sphere $S^5$.  Finally, it will be convenient to
only consider the traceless parts of $h_{\dot a\dot b}$ to avoid
mixing with other supergravity fields.%
\footnote{
  For example, the trace $h^{\dot a}_{\dot a}$ over $S^5$ indices
  mixes with fluctuations of the 5-form field strength
  $F_{\dot a \dot b \dot c \dot d \dot e}$ on $S^5$
  \cite{KRvN}.
}
Other than these constraints, the Kaluza-Klein mode may be
whatever one desires, and will determine, for example, the R
charge of the excitation.  In summary, we shall consider a 10-dimensional
field of the form
\begin {equation}
  h_{ab}(x,y) =
  \begin {cases}
     \phi(x) \, Y_{a b}(y) , & a,b \in \{{\mathsf{6,7,8,9,10}}\} ; \\
     0,                        & \mbox{otherwise},
  \end {cases}
\label {eq:h}
\end {equation}
where $x$ is the AdS$_5$-Schwarzschild coordinate, $y$ is the $S^5$
coordinate, $\phi$ is a 5-dimensional scalar field,
and $Y_{\dot a \dot b}(y)$ is any {\it traceless}\/
tensor spherical harmonic on $S^5$.
We will crudely summarize this choice by writing the field as
$h_{\dot a \dot b}$ when thinking of it as a 10-dimensional
field
and as $\phi$ when thinking of it as a 5-dimensional field.

Source operators dual to the fields (\ref{eq:h}) are those operators
of the form%
\footnote{
  See, for example, Table 7 of the review by D'Hoker and Freedman
  \cite{DFreview}.
  Here $X^k$ is shorthand for any symmetric product
  $X^{(i_1} X^{i_2} \cdots X^{i_k)}$ of $k$ factors of
  the three complex scalar
  fields $X^1$, $X^2$, $X^3$ of ${\cal N}{=}4$ SYM.
}
\begin {equation}
   O \sim \tr(\lambda\lambda\bar\lambda\bar\lambda X^k),
   \qquad
   k=0,1,2,\cdots
\label {eq:O}
\end {equation}
obtainable as supersymmetry descendants $Q^2 \bar Q^2$ of
$\tr(X^{k+4})$, where the $X$ are the adjoint-color scalar fields of
${\cal N}{=}4$ SYM and $\lambda$ are the gluinos.
The source operators (\ref{eq:O}) have conformal dimension
$\Delta=k+6$ and carry non-trivial R charge (and so the jets they
produce will carry R charge).
Different choices $k$ in (\ref{eq:O}) produce
different representations $(2,k,2)$ of the SU(4) R symmetry,
which correspond to different types of
tensor harmonics $Y_{\dot a \dot b}(y)$ in
(\ref{eq:h}).  None of the details of R charge representations,
which element we choose,
or tensor harmonics will matter for what follows.

It would be useful to check that our qualitative conclusions are not
sensitive to picking this particular class of source operators.
But, for most of this paper, we will simply stick to fields
(\ref{eq:h}) in the gravity
dual and so source operators (\ref{eq:O}).
However, in studying the leading corrections to
exponential tails in section \ref{sec:tail}, we will
make a slight generalization to also consider
the field $h_{\one\two}$, dual to the $T^{\one\two}$ component of
the stress tensor.

Throughout this paper, we will for simplicity
treat the conformal dimension $\Delta$ of
the source operator as parametrically of order one, $\Delta \sim 1$.
See ref.\ \cite{adsjet2}
for a discussion in the context of $\lambda{=}\infty$ stopping distances
of what happens when $\Delta \gg 1$ but with $\Delta$ still
parametrically small
compared to powers of $E$ and $\lambda$.


\section {The \boldmath$R^4$ correction}
\label {sec:R4}

We now want to discuss how higher-derivative corrections to the supergravity
action affect the 5-dimensional particle equations of motion.  It
will be helpful to do this first in the context of a concrete example of
a higher-derivative correction.
Though our final qualitative conclusions will not depend on the exact
pattern of higher-derivative supergravity corrections that arise from
Type IIB string theory, it will be convenient to start
the discussion with the first one that does,%
\footnote{
   For readers who are instead
   curious about what would happen in Gauss-Bonnet
   gravity, see Appendix
   \ref{sec:GB}.
}
which is $R^4$.


\subsection {\boldmath$R^4$ term in the 10-dimensional supergravity action}

The first corrections to low-energy supergravity arise from the
low-energy limit of the string-string scattering amplitude, for which
$\Nc{=}\infty$ corresponds to the tree-level amplitude.
The string-string scattering amplitude with 4 external graviton
states as legs gives rise to an $R^4$ interaction in 10-dimensional
supergravity,
where one may crudely think of each factor of $R$ as corresponding
to each leg.  Through use of the equations of motion, one may
in fact put the gravitational interaction in the form $C^4$ where $C$
is the Weyl tensor.  (We will not need to make such ``on-shell''
assumptions about higher-derivative corrections in our general analysis
later on.)  The details of how the four Weyl tensors are contracted
will not matter to our parametric analysis but, for the sake of
concreteness, here is the explicit modification to the
low-energy supergravity action:%
\footnote{
  Eq.\ (\ref{eq:C4BLS}) is nicely summarized in eqs.\ (3.1--3) of
  ref.\ \cite{BLS}
  and originates from refs.\ \cite{GrossWitten,C4}.  We have fixed the dilaton
  field $\varphi$ since we will not need to consider its fluctuations.
  (When comparing to more general results involving modular forms
  depending on $\varphi$, note
  that the $\Nc \to \infty$ limit corresponds to the large $\varphi$ limit.
  After taking the large $\varphi$ limit,
  one may rescale away any remaining dependence on the constant
  value of $\varphi$.)
}
\begin {equation}
   R \to R +
   \tfrac18 \, \zeta(3) \, \alpha'^3
   \left[
     C^{hmnk} C_{pmnq} {C_h}^{rsp} {C^q}_{rsk}
     +\tfrac12 C^{hkmn} C_{pqmn} {C_h}^{rsp} {C^q}_{rsk}
   \right]
\label {eq:C4BLS}
\end {equation}
or equivalently (via Bianchi identities)
\begin {equation}
   R \to R +
   \tfrac18 \, \zeta(3) \, \alpha'^3
   \left[
     C^{hmnk} C_{pmnq} {C_h}^{rsp} {C_{krs}}^q
     +\tfrac14 C^{hkmn} C_{pqmn} {C_h}^{prs} {C^q}_{krs}
   \right] ,
\label {eq:C4}
\end {equation}
where $R$ is the Ricci scalar in this equation, and
$\alpha'$ is the string tension.  The usual duality relation between
the string tension and 't Hooft coupling is \cite{MAGOO}
\begin {equation}
   \frac{\alpha'}{\Rads^2} = \lambda^{-1/2}
   \equiv (g_{\rm YM}^2 N_{\rm c})^{-1/2} ,
\end {equation}
where $\Rads$ is the $S^5$ radius.


\subsection{The $\phi$ equation of motion}

Let's now figure out the form of the corresponding linearized
equation of motion
for our 5-dimensional scalar field $\phi \sim h_{\dot a\dot b}$ in the
AdS$_5$-Schwarzschild background.  By ``linearized,'' we mean linearized
in $\phi$, not in the background metric.  That means that we want
terms in the action (\ref{eq:C4}) that are quadratic in $\phi$.

The $C^4$ correction term is suppressed by
$\alpha'^3 \propto \lambda^{-3/2} \ll 1$, and so the only way we can get
an unsuppressed correction is if there are compensating factors of
the large energy $E$ associated with the $\phi$ wave packet created
by the source.  The dominant correction will be the one with the most
powers of $E$.  Powers of $E$ will arise from derivatives
(with indices in AdS$_5$-Schwarzschild) hitting $\phi$.
So we should focus on the pieces of the $C^4$ term in the action
(\ref{eq:C4}) that are quadratic in $\phi$ and have as many
5-dimensional derivatives
acting on $\phi$ as possible.

One way to get a factor of $\phi \sim h_{\dot a\dot b}$
is to consider the piece of the
Weyl tensor $C_{ijkl}$ that involves two 5-dimensional
derivatives of the $S^5$ metric
fluctuation $h_{\dot a\dot b}$, e.g.
\begin {equation}
  C_{I\dot a J \dot b} \simeq -\tfrac12 \nabla_I \nabla_J h_{\dot a\dot b} .
\label {eq:CIaJb}
\end {equation}
Other terms, with fewer derivatives acting on $h_{\dot a\dot b}$, will
be suppressed because they do not generate as many factors of $E$.
Similarly, a term involving $S^5$ derivatives like
$\nabla_{\dot m} \nabla_{\dot n} h_{\dot a\dot b}$
arising from $C_{\dot m\dot a \dot n\dot b}$ will be suppressed compared
to (\ref{eq:CIaJb}) because $S^5$ derivatives of $\phi$ do not
yield factors of $E$.  The dominant terms in $C^4$ that contribute
to the linearized $\phi$ equation of motion will therefore
be those terms that have two factors of the form (\ref{eq:CIaJb})
and two factors of the background Weyl tensor
$\bar C_{ijkl}$ of (AdS$_5$-Schwarzschild)$\times S^5$.
Now use the fact that the background Weyl tensor
$\bar C_{ijkl}$ in this case vanishes unless all of its indices
live in AdS$_5$-Schwarzschild.%
\footnote{
  This is a statement about the uncorrected, i.e. $\lambda{=}\infty$,
  (AdS$_5$-Schwarzschild)$\times S^5$ background, and does not account
  for corrections to that background due to $\alpha'^3 C^4$.
  But, as we shall
  see, this is good enough for figuring out the leading correction
  to the $\phi$ equation of motion.
}
The dominant $C^4$ terms in the action then have the form
\begin {equation}
   \# \alpha'^3 (\nabla \nabla \phi) (\nabla \nabla \phi) \bar C \bar C ,
\label {eq:C4form}
\end {equation}
where $\#$ indicates some coefficient and
the suppressed indices on $\nabla$ and $\bar C$ are all five-dimensional
indices and contracted.

Note that such terms only arise from the first term in brackets
in (\ref{eq:C4}) and not the second.  For
$\tfrac14 C^{hkmn} C_{pqmn} {C_h}^{prs} {C^q}_{krs}$,
getting two factors of
the form (\ref{eq:CIaJb}) would require evaluating background Weyl
tensors $\bar C$ with at least one $S^5$ index, which gives zero.

The 5-dimensional
equation of motion for $\phi$ corresponding to (\ref{eq:C4form})
has the schematic form%
\footnote{
   For readers wishing to compare (i) the schematic equation
   (\ref{eq:eomform}) that summarizes the dominant terms in the
   high energy limit to (ii) the full, explicit equation of motion for
   a specific example (the bottom of the Kaluza-Klein tower for
   traceless $h_{\dot a\dot b}$), see Appendix \ref{app:dispersion}.
}
\begin {equation}
   [ \nabla \nabla
     + \# (\alpha')^3 \nabla\nabla \bar C \bar C \nabla \nabla ] \phi = 0 ,
\label {eq:eomform}
\end {equation}
where we have dropped the 5-dimensional mass term
(determined by
the conformal dimension of the source operator
and arising in part from $S^5$
derivatives on $\phi$ in the leading-order supergravity action).
As mentioned earlier, the mass is ignorable when computing the
stopping distance for $\ell_{\rm stop} \ll \ell_{\rm max}$.


\subsection{The WKB approximation}

Now let's look for wave solutions to the 5-dimensional equation of motion
(\ref{eq:eomform}) that have large, definite 4-momentum
$q_\mu$:
\begin {equation}
  \phi = \Phi(x^\five) \,  e^{i q_\alpha x^\alpha} .
\label {eq:phiq}
\end {equation}
(We will superpose such solutions to make wave packets in
the next section.)
Remember that Greek indices run over 4-dimensional space-time.
At high energy, we can make a WKB-like approximation and re-write
(\ref{eq:phiq}) as
\begin {equation}
  \phi = e^{i S(x^\five)} \,  e^{i q_\alpha x^\alpha}
\label {eq:phiWKB}
\end {equation}
where $S(x^\five)$ is large.
As discussed in refs.\ \cite{adsjet,adsjet2}, this approximation
only works sufficiently far from the boundary: $z \gg z_{\rm WKB}$ where
$z_{\rm WKB} \sim 1/\sqrt{-q^2}$.
But for $\ell_{\rm stop} \ll \ell_{\rm max}$
(i.e.\ $-q^2 \gg E^{2/3} T^{4/3}$)
that is good enough to analyze the stopping distance,
which is dominated by $z \sim z_\star$ given by (\ref{eq:zstar}).

Now use (\ref{eq:phiWKB}) in the equation of motion (\ref{eq:eomform}).
4-space derivatives $\nabla_\mu$ will give the largest contribution
when they hit the phase factor in (\ref{eq:phiWKB}) and bring down a
factor of the large $q_\alpha$ rather than hitting something involving
the background metric.  $x^\five$ derivatives $\nabla_\five$ will give
the largest contribution when they hit the $e^{i S(x^\five)}$ in
(\ref{eq:phiWKB}) and bring down a large factor of
$i \partial_\five S$ rather than
hitting something involving the background metric.  So, in the
large-energy limit, the dominant terms correspond to replacing
\begin {equation}
  \nabla_I \to i Q_I \equiv i (q_\mu, q_\five)
\label{eq:nabla}
\end {equation}
in the equation of motion for $\phi$, where
\begin {equation}
   q_\five \equiv \frac{\partial S}{\partial x^\five} \,.
\label {eq:q5def}
\end {equation}
We capitalize $Q_I$ just as a way of notationally emphasizing that it
is a 5-vector momentum.
The result of (\ref{eq:nabla}) is to replace the equation of motion
(\ref{eq:eomform}) by
\begin {equation}
   [ - Q Q
     + \# \alpha'^3 Q Q \bar C \bar C Q Q ] \phi = 0 .
\label {eq:eomQ}
\end {equation}
One may then solve this equation algebraically for $q_\five$ as
a function of $q_\mu$ and $x^\five$.  Integrating (\ref{eq:q5def})
then gives%
\footnote{
  There is a slight difference in sign convention between our work here
  and in some previous work by two of the authors.
  In ref.\ \cite{adsjet}, $q_\mu$ was the 4-momentum
  conjugate to the {\it boundary}\/ (rather than bulk) position,
  which would be $-q_\mu$ in the convention used in
  (\ref{eq:phiWKB}) and (\ref{eq:WKB}).
  With the convention used in our current paper, retarded
  boundary-to-bulk propagators (also known as
  advanced bulk-to-boundary propagators) correspond to solutions with
  $q_\five > 0$.
}
\begin {equation}
   \phi \simeq e^{iq_\alpha x^\alpha + i \int q_5 \> dx^5} ,
\label {eq:WKB}
\end {equation}
where $q_\five = q_\five(q_\mu,x^\five)$ solves (\ref{eq:eomQ}).

Just for the sake of being concrete, we give the explicit formula
for the equation of motion (\ref{eq:eomQ}) for the explicit
$C^4$ operator given by (\ref{eq:C4}):
\begin {equation}
   - (\bar g + \delta g)^{QS} Q_Q Q_S
   + \tfrac14 \, \zeta(3) \, \alpha'^3
     \overline{Q^H Q^K Q_P Q_Q {C_H}^{RSP} {C^Q}_{RSK}} \simeq 0,
\end {equation}
where overlines indicate that the 5-dimensional metric is taken to be
AdS$_5$-Schwarzschild (\ref{eq:metric}), and $\delta g$ is
the correction to that metric caused by the $C^4$ operator.

We do not need to work out $\delta g$.  It is enough to
note that this distortion $\delta g$ of the equilibrium metric
has nothing to do with the fact that we
have injected the $\phi$ wave packet into the system and so
is independent of $E$.  Its effects are
therefore suppressed by powers of $\lambda^{-1/2}$ with no compensating
factors of $E$.%
\footnote{
   One should pause to consider whether $\delta g$ might blow up faster
   than $\bar g$ at small $z$, since powers of $1/z$ do give powers of
   $E$ at the $z$ scale $z_\star$ of interest (\ref{eq:zstar}).
   However, as one takes $z \to 0$, AdS$_5$-Schwarzschild approaches
   AdS, and it is known that higher-derivative corrections do not modify
   the AdS solution \cite{KalloshRajaraman}.
   This means that the modifications to
   AdS$_5$-Schwarzschild will not become large as $z \to 0$.
}
So we will drop $\delta g$ and no longer
bother explicitly writing the overlines to indicate the
AdS$_5$-Schwarzschild metric:
\begin {equation}
   - Q^I Q_I
   + \tfrac14 \, \zeta(3) \, \alpha'^3
     Q^H Q^K Q_P Q_Q {C_H}^{RSP} {C^Q}_{RSK} \simeq 0.
\label {eq:QeomC4}
\end {equation}
We will discuss in a moment the parametric size of the $\alpha'^3$ correction
to the $\lambda{=}\infty$ dispersion relation $Q^I Q_I \simeq 0$.
But first, we will
briefly review how to turn a dispersion relation like (\ref{eq:QeomC4})
into a particle trajectory using the geometric optics approximation.


\subsection{The geometric optics (particle) approximation}

The geometric optics approximation consists of approximating the wave
packets as simultaneously having (i) well defined 5-momentum $Q_I$,
satisfying (\ref{eq:QeomC4}) above, and (ii) well defined
5-position $(x^\mu,x^\five)$.
One way to get the particle equation of motion is to start from
the WKB approximation (\ref{eq:WKB}) to the boundary-to-bulk
propagator and to make a wave packet by convolving with an
appropriate localized boundary source function
$\Lambda_L(x)$:
\begin {equation}
   \phi(x) \sim
   \int d^4q \>  e^{iq_\alpha x^\alpha + i \int q_\five(q,x^\five) \> dx^\five}
                           \tilde\Lambda_L(-q) .
\end {equation}
This integral can be done by saddle point methods, and the saddle point
condition is
\begin {equation}
   0 = \frac{\partial}{\partial q_\mu}
       \bigl[ iq_\alpha x^\alpha + i \int q_5(q,x^\five) \> dx^\five \bigr] ,
\end {equation}
which gives
\begin {equation}
   x^\mu = - \int dx^\five \> \frac{\partial q_\five}{\partial q_\mu} .
\label {eq:trajectory}
\end {equation}

Formally, we may then use this expression to find the generalization
\begin {equation}
   \ell_{\rm stop} \simeq
   - \int_0^{\zh} dz \> \frac{\partial q_\five}{\partial |\q|}
\label {eq:stopq5}
\end {equation}
of the stopping distance integral (\ref{eq:stop1}).
We will see later that this integral will require
care in interpretation in the region where $z$ is large enough that
the expansion in higher-derivative corrections breaks down,
as in fig.\ \ref{fig:fall2}.
Our focus will be on the relative importance
of higher-derivative corrections in the integrand at
$z \sim z_\star$.

See ref.\ \cite{adsjet2} for a $\lambda{=}\infty$
discussion of when the wave packet is small enough to treat as a particle.
The summary is that $L$ can be chosen appropriately so that
everything is fine at $z \sim z_\star$ when
$\ell_{\rm stop} \ll \ell_{\rm max}$.


\subsection{The relative importance of the \boldmath$C^4$ correction}
\label {sec:C4importance}

In cases where $C^4$ effects are a small correction, we may solve
the dispersion relation (\ref{eq:QeomC4}) iteratively.  That is,
first solve the $\lambda{=}\infty$ equation $Q^I Q_I = 0$ for
$q_\five$, and then plug that solution into the correction term
and solve
\begin {subequations}
\label{eq:Qeom2}
\begin {equation}
   Q^I Q_I
   = \tfrac14 \, \zeta(3) \, \alpha'^3
     Q^H Q^K Q_P Q_Q {C_H}^{RSP} {C^Q}_{RSK} \biggl|_{{\rm null}~Q_I}
\end {equation}
for $q_\five$.
Explicitly evaluating the AdS$_5$-Schwarzschild Weyl tensor,%
\footnote{
  $(C_{\mathsf{0101}},C_{\mathsf{1212}},C_{\mathsf{0505}},C_{\mathsf{1515}})
   = (f,1,-3,-f^{-1}) \times \Rads^2/\zh^4$,
  with all other components determined by symmetry.
}
one finds
\begin {equation}
   Q^H Q^K Q_P Q_Q {C_H}^{RSP} {C^Q}_{RSK} \biggl|_{{\rm null}~Q_I}
   = 24 \, \frac{z^{12} |\q|^4}{(\zh\Rads)^8} \,.
\label {eq:QQQQCC}
\end {equation}
\end {subequations}
Eq.\ (\ref{eq:Qeom2}), and the arguments leading up to it,
give the leading high-energy terms of the $C^4$-corrected dispersion relation.
Readers wishing to see the full
$C^4$-corrected equation of motion for comparison, without any
high-energy approximation, may find it in Appendix \ref{app:dispersion}.

Eq.\ (\ref{eq:Qeom2}) is already more detailed than we need for our
purposes---the specifics do not matter for our qualitative conclusions,
and it is easy to understand where the parametric dependence of
(\ref{eq:QQQQCC}) comes from without doing any explicit calculation,
at least for the case $z \ll \zh$ that will be of interest for
studying $z \sim z_\star$.
Giving a generic argument will help us to later discuss
yet-higher derivative corrections, where exact expressions for the
corrections are not generally known.

Schematically, the left-hand side of (\ref{eq:QQQQCC}) has the form
\begin {equation}
   g^{\bul\bul} g^{\bul\bul} g^{\bul\bul}
   g^{\bul\bul} g^{\bul\bul} g^{\bul\bul}
   Q_\bul Q_\bul Q_\bul Q_\bul
   C_{\bul\bul\bul\bul} C_{\bul\bul\bul\bul} ,
\end {equation}
where we will use bullets ($\scriptstyle{\bul}$)
to denote 5-dimensional indices
without focusing on the details of how they are contracted.
(i) The four powers of $Q_\bul$ produce four powers of $E$ (as long
as they are not contracted with each other).
(ii) The six powers of the inverse metric give six powers of
$z^2/\Rads^2$ for $z \ll \zh$, for a total of $z^{12}/\Rads^{12}$.
(iii) Finally, consider $C_{\bul\bul\bul\bul}$ for small $z$.
First think about pure AdS$_5$ space instead of AdS$_5$-Schwarzschild.
The Riemann tensor has size
\begin {equation}
   {R^{I}}_{JKL} \sim \frac{1}{z^2}
\end {equation}
by dimensional analysis (remembering that the overall normalization of
the metric cancels in ${R^{a}}_{bcd}$).
Lowering the upper index gives
\begin {equation}
   R_{IJKL} \sim \frac{\Rads^2}{z^4} .
\label {eq:RiemannAdS}
\end {equation}
Because pure AdS is a maximally symmetric space, the corresponding
Weyl tensor (which is the traceless part of Riemann) vanishes.
When we go from AdS$_5$ to AdS$_5$-Schwarzschild, the only change to the
metric is to replace $f=1$ by $f=1-(z/\zh)^4$, which introduces
corrections whose relative size at small $z$ is $(z/\zh)^4$.  So
(\ref{eq:RiemannAdS}) becomes
\begin {equation}
   R_{IJKL} \sim
   \frac{\Rads^2}{z^4} \left[1 + O\bigl(\frac{z^4}{\zh^4}\bigr)\right] .
\end {equation}
When we construct the Weyl tensor $C_{\alpha\beta\gamma\delta}$, we
know that the first term (the AdS one) will cancel and vanish.  So,
in AdS$_5$-Schwarzschild,
\begin {equation}
   C_{IJKL} \sim
   \frac{\Rads^2}{z^4} \times O\bigl(\frac{z^4}{\zh^4}\bigr)
   \sim \frac{\Rads^2}{\zh^4} .
\end {equation}
Now multiplying our considerations of (i) through (iii) above gives
\begin {equation}
   g^{\bul\bul} g^{\bul\bul} g^{\bul\bul}
   g^{\bul\bul} g^{\bul\bul} g^{\bul\bul}
   Q_\bul Q_\bul Q_\bul Q_\bul
   C_{\bul\bul\bul\bul} C_{\bul\bul\bul\bul}
   \sim
   \left( \frac{z^2}{\Rads^2} \right)^6
     \times E^4
     \times \left( \frac{\Rads^2}{\zh^4} \right)^2
   \sim
   \frac{z^{12} E^4}{(\zh \Rads)^8}
\end {equation}
for small $z$, consistent with the exact result (\ref{eq:QQQQCC}).

Now solving the 5-dimensional dispersion relation (\ref{eq:Qeom2})
for $q_\five(q_\mu,x^\five)$ gives
\begin {equation}
   q_5
   \simeq
   \sqrt{
       g_{55}
       \left(
          - q_\mu g^{\mu\nu} q_\nu
          + \frac{\coeff z^{12} |\q|^4}{\zh^8 \Rads^2}
       \right)
   }
   ,
\label {eq:q5C4}
\end {equation}
where
\begin {equation}
   \coeff
   \equiv \frac{24}{\Rads^6} \times \tfrac14 \, \zeta(3) \,\alpha'^3
   = \frac{6 \, \zeta(3)}{\lambda^{3/2}}
\label {eq:epsilon}
\end {equation}
is small.

At this point, we could measure the parametric importance of the
$C^4$ correction simply by comparing the relative sizes of
the $\coeff z^{12} |\q|^4/\zh^8 \Rads^2$ and
$-q_\mu g^{\mu\nu} q_\nu$ terms in (\ref{eq:q5C4}) at $z \sim z_\star$.
But, for the sake of being slightly more explicit, let's first
use (\ref{eq:q5C4}) to get the stopping distance integral
(\ref{eq:stopq5}):
\begin {equation}
   \ell_{\rm stop} \simeq
   \int_0^{\zh} dz \>
   \frac{
     |\q| \left[ 1 - \frac{2 \coeff z^{10}}{\zh^8} |\q|^2 \right]
   }{
     \sqrt{
          - q^2 + \frac{z^4}{\zh^4} |\q|^2
          + \frac{\coeff z^{10}}{\zh^8} |\q|^4 f
     }
   } ,
\label {eq:stopC4}
\end {equation}
where $q^2 \equiv q_\mu \eta^{\mu\nu} q_\nu$
denotes the 4-momentum virtuality.
The $\lambda{=}\infty$ result (\ref{eq:stop1}) corresponds to
$\coeff = 0$.  There are various features of the integrand
in (\ref{eq:stopC4}) that need to be discussed, but first let's
look at the relative size of the $C^4$ correction at
$z_\star$ (\ref{eq:zstar}).  Under the square root in the denominator,
the $-q^2$ and $z^4 |\q|^2/\zh^4$ terms are the same size at
$z \sim z_\star$ --- that's how $z_\star$ was determined in the first
place.  Since $z_\star \ll \zh$, we have $f \simeq 1$, and the relative
size of the correction term is
\begin {subequations}
\label {eq:importanceC4}
\begin {equation}
  \mbox{Importance($C^4$)}
  \sim
  \frac{ \frac{\coeff z_\star^{10}}{\zh^8} E^4 }{ -q^2 }
  \sim
  \frac{(-q^2)^{3/2}}{\lambda^{3/2} E T^2} \,.
\label {eq:importanceC4q2}
\end {equation}
Using (\ref{eq:lmax}) and (\ref{eq:stop}), this may be rewritten as
\begin {equation}
  \mbox{Importance($C^4$)}
  \sim
  \left( \frac{ \lambda^{-1/4} \ell_{\rm max} }{ \ell_{\rm stop} } \right)^6
  \sim
  \lambda^{-1/2}
  \left( \frac{ \lambda^{-1/6} \ell_{\rm max} }{ \ell_{\rm stop} } \right)^6
  ,
\end {equation}
\end {subequations}
which gives the $R^4$ (red) line in fig.\ \ref{fig:corrections}.

The numerator correction in (\ref{eq:stopC4}) is less important at
$z \sim z_\star$.  The relative size of its correction to the
$\lambda{=}\infty$ integrand is
\begin {equation}
  \frac{ \frac{\coeff z^{10}}{\zh^8} E^2 }{ 1 } \,
\end {equation}
which is smaller than (\ref{eq:importanceC4q2}) at $z\sim z_\star$
by a factor of $-q^2/E^2 \ll 1$.
One seemingly disturbing feature of the numerator correction is its sign
for large enough $z$.  For
\begin {equation}
   z \gg \left( \frac{\lambda^{3/4} T}{E} \right)^{1/5} \zh
\label {eq:C4problem}
\end {equation}
(which is much larger than $z_\star$)
the integrand is large and negative.
If the integral
(\ref{eq:stopC4}) is blindly integrated up to $\zh$ as written,
one would find a rather extreme case of fig.\ \ref{fig:fall2}c.
However, we show in sections \ref{sec:zbad} and Appendix \ref{app:C4}
that the expansion in higher-derivative
corrections breaks down
well before one reaches $z$'s
as large as (\ref{eq:C4problem}).  And so the situation,
for stopping distances in the safe region
$\ell_{\rm stop} \gg \lambda^{-1/6} \ell_{\rm max}$ of
fig.\ \ref{fig:corrections}, is that of fig.\ \ref{fig:fall2}a.
Following our assumptions from section \ref{sec:importance}, we
therefore stick to (\ref{eq:importanceC4}) as the measure of the
importance of $C^4$ corrections.

As far as $C^4$ corrections are concerned,
our main results in this paper
(fig.\ \ref{fig:corrections}) only require the parametric
information (\ref{eq:importanceC4}) on the importance of the
$C^4$ correction at $z \sim z_\star$.
One might be tempted to attempt to extract an exact size for
the leading correction from the explicit integral
(\ref{eq:stopC4}).  We show in Appendix \ref{app:C4} why
this will fail.
Eq.\ (\ref{eq:importanceC4}) is adequate for determining whether or not
the correction will be small, given our assumptions in
section \ref{sec:importance}.  But the appendix shows that,
even when the correction is small,
the integrand cannot be trusted for the range of $z$ values
required to use (\ref{eq:importanceC4}) for a precise calculation
of the correction.


\section {The \boldmath$D^{2n} R^4$ corrections}
\label{sec:D2nR4}

In this section, we'll investigate our first sequence of
higher and higher derivative corrections to the 10-dimensional supergravity
dual, by looking at $R^4$ terms with higher and higher powers of
covariant derivatives.
(We'll save adding extra powers of the Riemann curvature
for the next section.)  The parametric size of such corrections
can be studied in complete generality, without focusing on precise
formulas for the terms $D^{2n}R^4$.
However, for the sake of being
concrete and making contact with standard string theory results,
we'll first pause to briefly review what's known about
these operators in the case of duality with Type II string theory.


\subsection{Review: 4-point string amplitude}

Just like the $R^4$ interaction in supergravity arises from the low-energy
limit of graviton-graviton scattering in string theory, the $D^{2n}R^4$
operators arise by looking more generally at the energy/momentum
dependence of that scattering.  At tree level (appropriate for
$\Nc{=}\infty$), the energy dependence of the amplitude is
captured by an overall factor
\begin {equation}
  T(s,t,u) =
  -
  \frac{
     \Gamma(-\alpha' s/4) \, \Gamma(-\alpha'  t/4) \, \Gamma(-\alpha' u/4)
  }{
     \Gamma(1+\alpha' s/4) \, \Gamma(1+\alpha' t/4) \, \Gamma(1+\alpha' u/4)
  }
  \,,
\label {eq:T}
\end {equation}
where $s$, $t$, and $u$ are the Mandelstam variables (in 10 dimensions).
This result is an ``on-shell'' result, which means it is derived for
string scattering in a flat-space background with the external momenta
on-shell.  That is, the result assumes $q_a q^a = 0$ for each of the four
10-dimensional external momenta, which means $s+t+u=0$.
Expanding $T(s,t,u)$  in powers of momenta gives \cite{GreenVanhove}
\begin {align}
   T &= \frac{64}{{\alpha'}^3 s t u} \,
   \exp\left[
     \sum_{n=1}^\infty \frac{2\,\zeta(2n+1)}{2n+1}
     \left(\frac{\alpha'}{4}\right)^{2n+1}
     (s^{2n+1} + t^{2n+1} + u^{2n+1})
   \right]
\nonumber\\
   &= \frac{3}{\sigma_3} + 2\,\zeta(3)
      +\zeta(5) \, \sigma_2
      +\tfrac23 \, \zeta^2(3) \, \sigma_3
      +\tfrac12 \, \zeta(7)\,(\sigma_2)^2
      +\tfrac23 \, \zeta(3)\, \zeta(5) \, \sigma_2 \sigma_3
      +\cdots ,
\label {eq:Texp}
\end {align}
where
\begin {equation}
   \sigma_k \equiv \left( \frac{\alpha'}{4} \right)^k (s^k+t^k+u^k) .
\end {equation}
The first term in the expansion (\ref{eq:Texp})
corresponds to scattering that occurs through the interchange
of an intermediate graviton, and this process is already accounted for
by the usual Einstein-Hilbert piece $R$ of the low-energy supergravity action.
The second term in (\ref{eq:Texp}), when generalized to curved space,
gives the $R^4$ interaction previously discussed in
section \ref{sec:R4}.  In order, the remaining terms give interactions that
are schematically of the form $D^4 R^4$, $D^6 R^4$, etc.

In what follows, we will look at the parametric size of the effects of
these interactions.
Readers may wonder why we should think about a
derivative expansion $D^{2n} R^4$, and worry about where that expansion
breaks down, when we already know that the expansion sums up to
(\ref{eq:T}).  There are two reasons.  First, it will turn out that
in those cases where the expansion is breaking down it will also be
the case that the ``on-shell'' assumption $q^a q_a = 0$ for the
graviton momenta will also break down.  So it is safest to not make any
explicit assumptions about the detailed form of the $D^{2n} R^4$
interactions.  Secondly, analyzing the derivative expansion will provide
a useful warm-up to more generally analyzing higher-derivative corrections
$D^{2n} R^m$, which, as shown in fig.\ \ref{fig:corrections}, are
equally important when the derivative expansion breaks down for
$D^{2n} R^4$.  As we will see, everything goes wrong at the same time,
and so the explicit formula (\ref{eq:T}) for the 4-point amplitude
does not seem useful then.


\subsection{Factors of \boldmath$\alpha' Q Q$}
\label {sec:D2C4QQ}

In section \ref{sec:R4},\
we examined the effects of an $\alpha'^3 C^4$ interaction
and found a 5-dimensional dispersion relation for the
linearized scalar field $\phi$ with schematic form
\begin {equation}
   Q^I Q_I =
   \alpha'^3
   g^{\bul\bul} g^{\bul\bul} g^{\bul\bul}
   g^{\bul\bul} g^{\bul\bul} g^{\bul\bul}
   Q_\bul Q_\bul Q_\bul Q_\bul
   C_{\bul\bul\bul\bul} C_{\bul\bul\bul\bul}
   \biggr|_{{\rm null}~Q_I} .
\label {eq:C4dispRecap}
\end {equation}
This arose from terms in the Lagrangian quadratic in $\phi$,
with 5-dimensional form
\begin {equation}
   \alpha'^3 (\nabla\nabla\phi)(\nabla\nabla\phi) C C
\label {eq:ppCC}
\end {equation}
coming from the 10-dimensional $\alpha'^3 C^4$.

Forget about string theory expressions and just think about
what would happen if we went from $\alpha'^3 C^4$ to something of the form
$\alpha'^4 D^2 C^4$.
(Note that the additional factor of $\alpha'$ that accompanies the $D^2$
is simply a consequence of dimensional analysis.)
Naively, we might think that the largest
contribution arises from the case where both of the new derivatives have
5-dimensional indices and hit $\phi$'s, modifying the right-hand-side
of (\ref{eq:C4dispRecap}) to include an additional factor of
\begin {equation}
   \alpha' g^{\bul\bul} Q_\bul Q_\bul .
\label {eq:alphaD2fac}
\end {equation}
Here the new indices might contract with the other indices in
(\ref{eq:C4dispRecap}) or with each other.
If we
then note that $Q_\mu$ grows like $E$, we might at first guess that
the parametric size of the additional factor (\ref{eq:alphaD2fac})
could be as large as
\begin {equation}
   \alpha' \times \frac{z^2}{\Rads^2} \times E \times E ,
\label {eq:alphaD2fac2}
\end {equation}
but this is an overestimate.  If all six $Q$'s in
\begin {equation}
   \alpha'^4
   g^{\bul\bul} g^{\bul\bul} g^{\bul\bul}
   g^{\bul\bul} g^{\bul\bul} g^{\bul\bul}
   g^{\bul\bul}
   Q_\bul Q_\bul
   Q_\bul Q_\bul Q_\bul Q_\bul
   C_{\bul\bul\bul\bul} C_{\bul\bul\bul\bul}
\end {equation}
are contracted with indices of the two Weyl tensor factors, the
result must vanish because $C_{IJKL}$ is anti-symmetric in
$IJ$ and $KL$.  As a result, two of the $Q$'s must contract with
each other, and so the cost of the factor (\ref{eq:alphaD2fac})
is
\begin {equation}
   \alpha' Q^I Q_I
\end {equation}
instead of (\ref{eq:alphaD2fac2}).  In the $\lambda{=}\infty$
calculation, $Q^I Q_I = 0$.  In our calculation here, the
effects discussed earlier arising from the $C^4$ correction
change this to (\ref{eq:Qeom2}),
\begin {equation}
   Q^I Q_I \sim
   \alpha'^3 \, \frac{z^{12} E^4}{(\zh\Rads)^8} \,.
\label {eq:QQ}
\end {equation}
So the size of the factor (\ref{eq:alphaD2fac}) at $z\sim z_\star$
is
\begin {equation}
   \alpha' Q^I Q_I \bigl|_{z\sim z_\star}
   \sim
   \alpha'^4 \, \frac{z_\star^{12} E^4}{(\zh\Rads)^8}
   \sim
   \frac{(-q^2)^3}{\lambda^2 E^2 T^4}
   \sim
   \left( \frac{\lambda^{-1/6} \ell_{\rm max}}{\ell_{\rm stop}} \right)^{12}
   .
\label {eq:alphaD2fac3}
\end {equation}

As we will discuss, (\ref{eq:alphaD2fac}) is not, in fact, the dominant
contribution for large $\ell_{\rm stop}$
simply because of the suppression from having to contract
the $Q$'s.  But let's focus on this type of contribution for a
moment longer.  First, (\ref{eq:alphaD2fac3}) tells us that this
particular contribution from $\alpha'^4 D^2 C^4$ becomes just as
important as $\alpha'^3 C^4$ when
$\ell_{\rm stop} \sim \lambda^{-1/6} \ell_{\rm max}$, and so this
is our first example of the breakdown of the
expansion in higher-derivative corrections depicted in
Fig.\ \ref{fig:corrections}.  If we add yet another factor of
$\alpha' D^2$ to go to $\alpha'^5 D^4 C^4$, and consider just
the contributions of the form (\ref{eq:alphaD2fac}) for that
factor as well, then we will get another factor of
(\ref{eq:alphaD2fac3}), which will also not be suppressed at
$\ell_{\rm stop} \lesssim \lambda^{-1/6} \ell_{\rm max}$.
Finally, note that all of the effects discussed so far are
arising from $\alpha' Q^I Q_I$ factors, which is just the
dominant piece of 10-dimensional $\alpha' q^a q_a$ factors.
These are precisely the sort of factors that are left out
of standard string theory ``on-shell'' results for
higher-derivative corrections $D^{2n} R^4$, but they become important for
$\ell_{\rm stop} \lesssim \lambda^{-1/6} \ell_{\rm max}$.


\subsection{The dominant factors}
\label {sec:D2cost}

If we add a factor of $\alpha' D^2$ and neither derivative hits a
$\phi$, then there will be no powers of $E$ to compensate the
suppression from $\alpha'$.  The dominant terms come from the case
where one derivative hits a $\phi$ and the other hits the background
field:
\begin {equation}
   \alpha'^4
   g^{\bul\bul} g^{\bul\bul} g^{\bul\bul}
   g^{\bul\bul} g^{\bul\bul} g^{\bul\bul}
   g^{\bul\bul}
   Q_\bul
   Q_\bul Q_\bul Q_\bul Q_\bul
   D_\bul C_{\bul\bul\bul\bul} C_{\bul\bul\bul\bul} \,.
\end {equation}
Since the background field depends only on $x^\five$,
it is natural to consider the case where the derivative $D_\bul$
hitting the background Weyl tensor is a $D_\five$.
We will address this (relatively straightforward) case here.
The contribution of terms involving other components
$D_\mu$ of $D_\bul$ hitting the background is slightly more
subtle and will be left to Appendix \ref{app:details}.

Having the $D_\bul$ which hits the background Weyl tensor be
$D_\five$ means
(by 4-dimensional parity invariance) that one of the
$Q_\bul$'s must be $q_\five$.  Parametrically, a $D_5$ on the background
Weyl tensor has size $z^{-1}$ for $z \ll \zh$
(such as $z \sim z_\star$).
So the factor of $\alpha' D^2$ has cost
\begin {equation}
   \alpha' g^{\five\five} q_\five \, D_\five(\mbox{on bkgd})
   \sim
   \alpha' \times \frac{z^2}{\Rads^2} \times q_\five \times z^{-1} .
\label {eq:D2cost1}
\end {equation}
If you think of the factors in (\ref{eq:ppCC}) as
a 4-point amplitude with the two $\phi$ factors being legs
1 and 3 and the two $C$ factors being legs 2 and 4, then
the cost shown above corresponds to (the curved background
generalization of) an $\alpha' s$ or $\alpha' u$ factor in
the string amplitude expansion (\ref{eq:Texp}).

For the size of $q_\five$, we can just take the $\lambda{=}\infty$
result from $Q^I Q_I \simeq 0$,
\begin {equation}
   q_\five \simeq \sqrt{g_{55} (-q_\mu g^{\mu\nu} q_\nu)}
   \simeq \sqrt{-q^2 + \frac{z^4}{\zh^4} |\q|^2} .
\label {eq:qfive1}
\end {equation}
This is just (\ref{eq:q5C4}) with $\coeff$ ignored.
At $z \sim z_\star$, the two terms under the square root in (\ref{eq:qfive1})
have comparable size, and so
\begin {equation}
  q_\five \bigl|_{z \sim z_\star} \sim \sqrt{-q^2} .
\end {equation}
The cost (\ref{eq:D2cost1}) of $\alpha' D^2$ is then
\begin {equation}
   \alpha' D^2 |_{z \sim z_\star}
   \sim
   \frac{\alpha' z_\star \sqrt{-q^2}}{\Rads^2}
   \sim
   \frac{(-q^2)^{3/4}}{\lambda^{1/2} E^{1/2} T}
   \sim
   \left( \frac{\lambda^{-1/6} \ell_{\rm max}}{\ell_{\rm stop}} \right)^3
   .
\label {eq:D2cost2}
\end {equation}
This indeed dominates over (\ref{eq:alphaD2fac3}) in the regime
$\ell_{\rm stop} \gg \lambda^{-1/6} \ell_{\rm max}$ where the
expansion in higher-derivative corrections has not already broken down.
Multiplying the importance (\ref{eq:importanceC4}) of $C^4$
by any number of factors (\ref{eq:D2cost2}) gives
\begin {equation}
  \mbox{Importance($D^{2n} C^4$)}
  \sim
  \lambda^{-1/2}
  \left( \frac{ \lambda^{-1/6} \ell_{\rm max} }{ \ell_{\rm stop} }
       \right)^{3n+6} ,
\end {equation}
which is shown by the $D^{2n} R^4$ curves in fig.\
\ref{fig:corrections}.  Here and through most of this paper, we will be
a little sloppy in distinguishing $R^m$ and $C^m$ in our discussion.
We address that sloppiness in
appendix \ref{sec:R0} and show that it makes no
difference for our qualitative conclusions.


\subsection {Beyond supergravity fields?}

Throughout this paper, in our treatment of the gravity dual,
we focus on the dynamics of standard supergravity fields
in AdS$_5$-Schwarzschild.  That is, we do not explicitly consider
excited string states, which would correspond to a tower of
extremely-massive fields in the supergravity theory, with
masses of order $1/\sqrt{\alpha'}$.  In our analysis, the effects
of such excitations only appear indirectly through their effects
on the effective interactions (such as $R^4$) between low-mass
fields---that is, their only effects are as possible intermediate
states in the string-scattering amplitudes which determine those
effective interactions.  One might wonder whether, for jets of
sufficiently high energy $E$, it is possible to instead directly
excite such high-mass states ``on-shell'' in the gravity picture.
If so, we would need to explicitly include such fields in our analysis.

However, on-shell string excitations can be ignored precisely when the
derivative expansion is under control, i.e.\ when
$\ell_{\rm stop} \gg \lambda^{-1/6} \ell_{\rm max}$
in fig.\ \ref{fig:corrections}.  To see
this, consider, for example, the string-string scattering amplitude
factor (\ref{eq:T}).  This expression has poles when $\alpha's/4$ (or
similarly $\alpha' t/4$ or $\alpha' u/4$) is a non-negative integer,
corresponding
to an on-shell intermediate string state.  To excite a massive string
state therefore requires $s$ large enough that
$\alpha' s/4 \ge 1$ (or similar for $\alpha' t/4$ or $\alpha' u/4$).
But we've just seen in the previous sub-sections that
$\alpha' s$ and $\alpha't$ and $\alpha' u$ are all parametrically
{\it small}\/ when $\ell_{\rm stop} \gg \lambda^{-1/6} \ell_{\rm max}$.
That is, the condition that the derivative expansion be under control
is the same as the condition that the 5-dimensional Mandelstam
variables are too small to create on-shell massive string excitations.


\section {Higher powers of curvature}
\label {sec:Rm}

\subsection {\boldmath{$C^5$}}

Now consider what happens if we include an additional factor of the Weyl
tensor and look at an $\alpha'^4 C^5$ term in the 10-dimensional
Lagrangian.  There is, in fact, no independent
$D^{2n} C^5$ term with $n{=}0$ in
Type II supergravity \cite{Stieberger},
but the $n{=}0$ case will make for a useful warm-up.  The
additional power of $\alpha'$ that accompanies the additional power of
$C$ is a consequence of dimensional analysis.  In 5-dimensional
language, the terms quadratic in the scalar field $\phi$ will have the
form
\begin {equation}
   \alpha'^4 (\nabla\nabla\phi)(\nabla\nabla\phi)\bar C\bar C\bar C
\end {equation}
So all we've done by going from $\alpha'^3 C^4$ to
$\alpha'^4 C^5$ is to modify the right-hand side of
of the $C^4$ dispersion relation
(\ref{eq:C4dispRecap}) to include an additional factor of
\begin {equation}
  \alpha' g^{\bul\bul} g^{\bul\bul} C_{\bul\bul\bul\bul} \,.
\label{eq:C5factor}
\end {equation}
[Since the Weyl tensor is traceless, the additional indices necessarily
contract with the other indices in (\ref{eq:C4dispRecap}).]
By the same parametric counting as in section \ref{sec:C4importance},
this new factor is (for $z \ll \zh$) of order
\begin {equation}
  \alpha' g^{\bul\bul} g^{\bul\bul} C_{\bul\bul\bul\bul} \sim
  \alpha' \times \left(\frac{z^2}{\Rads^2}\right)^2
          \times \frac{\Rads^2}{\zh^4}
  \sim \lambda^{-1/2} \, \frac{z^4}{\zh^4} \ll 1 .
\end {equation}
There and no additional powers of energy to compensate the factor
of $\lambda^{-1/2}$, and so a $C^5$ correction, if there were one,
would always be small compared to a $C^4$ correction.


\subsection {\boldmath{$D^2 C^5$}}

Now add two more derivatives and instead consider $\alpha'^5 D^2 C^5$.
Naively, the dominant term is the one where both derivatives hit $\phi$
and give powers of $Q_I$ so that the net cost of adding factors
$\alpha'^2 D^2 C$ to the original $\alpha'^3 C^4$ is
\begin {equation}
  \alpha'^2 g^{\bul\bul} g^{\bul\bul} g^{\bul\bul}
  Q_\bul Q_\bul C_{\bul\bul\bul\bul} \,.
\label {eq:D2C5factor}
\end {equation}
Unlike what happened for $D^2 C^4$ in section \ref{sec:D2C4QQ}, here
the naive reasoning is correct because the two new factors of $Q$
need not contract with each other---they can instead contract with
indices of the new factor of $C$ in the combination
$Q^I Q^K C_{IJKL}$.  The parametric size of (\ref{eq:D2C5factor})
is therefore simply
\begin {equation}
  \alpha'^2 \times \left(\frac{z^2}{\Rads^2}\right)^3
        \times E^2 \times \frac{\Rads^2}{\zh^4}
  \sim \frac{z^6 E^2}{\lambda\zh^4} \,.
\label {eq:D2C5factor1}
\end {equation}
At $z \sim z_\star$, this is
\begin {equation}
  \frac{z_\star^6 E^2}{\lambda\zh^4}
  \sim \frac{(-q^2)^{3/2}}{\lambda  E T^2}
   \sim
   \left( \frac{\lambda^{-1/6} \ell_{\rm max}}{\ell_{\rm stop}} \right)^6
   .
\label {eq:D2C5factor2}
\end {equation}
And so $D^2 C^5$ becomes just as important as $C^4$ when
$\ell_{\rm stop} \sim \lambda^{-1/6} \ell_{\rm max}$, as in
fig.\ \ref{fig:corrections}.


\subsection {\boldmath{$D^{2n} C^{m}$}}

Every time we add another factor of $C$, we can also add
a pair of large derivatives, just like above.
For each new factor of $\alpha'^2 D^2 C$ added,
the cost is another factor of (\ref{eq:D2C5factor2}).  Multiplying
$k$ such factors times the importance (\ref{eq:importanceC4}) of
$C^4$ then gives
\begin {equation}
  \mbox{Importance($D^{2k} C^{4+k}$)}
  \sim
  \lambda^{-1/2}
  \left( \frac{ \lambda^{-1/6} \ell_{\rm max} }{ \ell_{\rm stop} }
       \right)^{6k+6} ,
\end {equation}
which is depicted, for example, by the $D^2 R^5$ and $D^4 R^6$ lines
in fig.\ \ref{fig:corrections}.

Starting from $D^{2k} C^{4+k}$, we may then add further derivatives,
but the parametric cost for additional derivatives
will be the same as the discussion in section \ref{sec:D2cost},
so that
\begin {equation}
  \mbox{Importance($D^{2k+2n} C^{4+k}$)}
  \sim
  \lambda^{-1/2}
  \left( \frac{ \lambda^{-1/6} \ell_{\rm max} }{ \ell_{\rm stop} }
       \right)^{6k+3n+6} ,
\end {equation}
also depicted in fig.\ \ref{fig:corrections}.
The upshot is that, when considering higher-derivative terms
$A$ in the supergravity Lagrangian,
the subset with the dominant
effect for a given engineering dimension $\operatorname{dim}A$
has importance
\begin {equation}
  \lambda^{-1/2}
  \left( \frac{ \lambda^{-1/6} \ell_{\rm max} }{ \ell_{\rm stop} }
       \right)^{\frac32 \operatorname{dim}A - 6} .
\end {equation}
But we will need to finish our arguments, and consider other possible
corrections, in order to verify this.


\subsection{Remaining details}

We've now seen the basic structure of corrections that give rise
to fig.\ \ref{fig:corrections}, but there are still a few details
to clear up.  So far, we have considered only powers of the Weyl
curvature tensor.  In appendix \ref{app:details}, we show that it
will not matter, qualitatively, if supergravity interactions instead involved
the full Riemann tensor at some order in derivatives.  Along the way,
we also show that it is unimportant for the qualitative conclusions
about the expansion that the first correction
was $C^4$ rather than some lower power like $C^2$ and $C^3$.
The appendix also addresses the effects of supergravity interactions
involving the 5-form field strength $F$, which has a non-vanishing
background value in (AdS$_5$-Schwarzschild)$\times S^5$.


\section {What happens for \boldmath$z \gg z_\star$?}
\label {sec:zbad}

By making the reasonable assumptions that we outlined
in section \ref{sec:importance}, we have managed to analyze
the question of when corrections become important by focusing
on particle trajectories at $z \sim z_\star$.  As $z$ increases
beyond this scale, the forward progress of the trajectory slows
to a stop as in fig.\ \ref{fig:fall2}a.  We previously asserted
that at some scale $z_{\rm bad} \gg z_\star$ the expansion in
higher-derivative corrections would eventually break down, as also
depicted in fig.\ \ref{fig:fall2}a, even if
the expansion was well behaved at $z_\star$.  Here we will take a moment
to identify the size of $z_{\rm bad}$.

Start by considering the cost of an $\alpha' D^2$ factor, which we
analyzed in section \ref{sec:D2cost} for
$z \sim z_{\star}$.
At larger $z$, with $z_\star \lesssim z \ll \zh$,
(\ref{eq:qfive1}) gives
\begin {equation}
  q_\five \bigr|_{z \gtrsim z_\star} \sim
  \frac{z^2 E}{\zh^2} .
\end {equation}
Then the cost (\ref{eq:D2cost1}) of each $\alpha' D^2$ factor is
\begin {equation}
   \alpha' D^2 |_{z \gtrsim z_{\star}}
   \sim
   \frac{\alpha' z q_\five}{\Rads^2}
   \sim
   \frac{z^3 E T^2}{\lambda^{1/2}} \,.
\end {equation}
This cost is unsuppressed for $z \gtrsim z_{\rm bad}$, where
\begin {equation}
   z_{\rm bad} \sim \frac{\lambda^{1/6}}{E^{1/3} T^{2/3}} \,.
\label {eq:zbad}
\end {equation}

The same constraint arises from the other important corrections that
we analyzed.  For instance, the cost (\ref{eq:D2C5factor1}) of adding an
$\alpha'^2 D^2 C$ factor was $z^6 E^2/\lambda \zh^4$, which
also becomes unsuppressed at the same $z \gtrsim z_{\rm bad}$.

Note that the requirement
$\ell_{\rm stop} \gg \lambda^{-1/6} \ell_{\rm max}$
for the expansion in corrections to be well-behaved in
fig.\ \ref{fig:corrections} is the same condition as requiring
$z_{\rm bad} \gg z_\star$.




\section {Exponential Tails}
\label {sec:tail}

So far our discussion of corrections has been parametric and
therefore qualitative in nature.  Our analysis has also relied
on the geometric optics approximation and so only been valid
for $\ell_{\rm stop} \ll \ell_{\rm max}$, though we then
parametrically extrapolated our results to
$\ell_{\rm stop} \sim \ell_{\rm max}$.
We have seen that the expansion in higher-derivative corrections
should be well-behaved at $\ell_{\rm stop} \sim \ell_{\rm max}$,
and the dominant correction is the $C^4$ term in the effective
supergravity action.  In this section, we will make an explicit calculation
of the effect of this correction on a quantity related to
the maximum stopping distance.


\subsection {The exponential tail scale \boldmath$\ell_{\rm tail}$}

\begin {figure}
\begin {center}
  \includegraphics[scale=0.3]{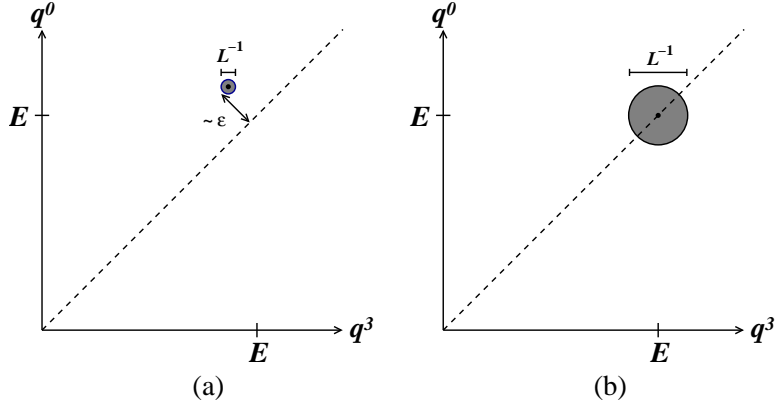}
  \caption{
     \label{fig:source}
     Qualitative picture of
     momenta contributing to the source (\ref{eq:source})
     used to generate jets
     (a) for the case $L^{-1} \ll \epsilon \ll E$ of (\ref{eq:bark2}),
     where one may use geometric optics, and (b) the case
     $\epsilon = 0$, which has components with arbitrarily small
     $-q^2$ and which can generate jets with the largest stopping
     distances.
  }
\end {center}
\end {figure}

\begin {figure}
\begin {center}
  \includegraphics[scale=0.4]{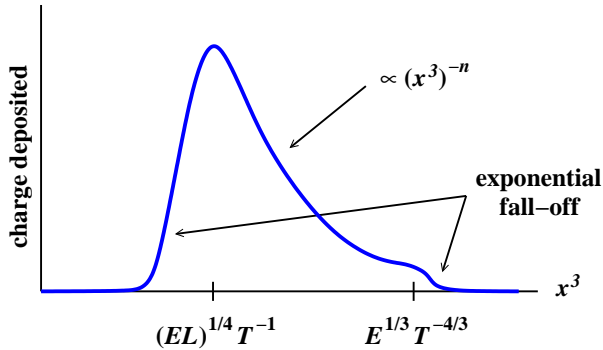}
  \caption{
     \label{fig:stop}
     The average deposition of charge as a function of $x^\three$ for
     jets created by source momenta of the form of fig.\
     \ref{fig:source}b.  The power $n$ of the algebraic fall-off
     for $(EL)^{1/4} T^{-1} \ll x^\three \ll \ell_{\rm max}$ depends on
     the source operator \cite{adsjet2}.
  }
\end {center}
\end {figure}

We first need a crisp definition of the ``maximum'' stopping distance,
or of something related to it.  To that end, consider the sources
that we have used so far in this paper.  The source
operator has been weighted with
\begin {equation}
   e^{i\bar k\cdot x} \, \Lambda_L(x)
\label {eq:sourcefn}
\end {equation}
as in (\ref{eq:source}), where the envelope function
$\Lambda_L(x)$ is localized
within a distance $L$ in both space and time, and where
\begin {equation}
   \bar k = (E+\epsilon,0,0,E-\epsilon)
   \qquad \mbox{with} \qquad
   \frac{1}{L} \ll \epsilon \ll E
\label {eq:bark2}
\end {equation}
as in (\ref{eq:koffshell}).  The 4-virtuality $-q^2$
of the source is
$\simeq -\bar k^2 \simeq 4\epsilon E$.  The support for
the source function (\ref{eq:sourcefn}) in momentum space is
depicted qualitatively in fig.\ \ref{fig:source}a
for the case (\ref{eq:bark2}).
As we make
$\epsilon$ (and so $-q^2$) smaller and smaller, the momenta move
closer and closer to the line $q^0=q^3$ and the stopping distance
becomes longer and longer.  To include virtualities as small
as possible, and so the longest possible stopping distances,
we may instead just take a momentum distribution that straddles
the $q^0 = q^3$ line, as in fig.\ \ref{fig:source}b.  This just
corresponds to (\ref{eq:sourcefn}) with
\begin {equation}
   \bar k = (E,0,0,E)
\label {eq:nullkbar}
\end {equation}
instead of (\ref{eq:bark2}).
One should choose $L \ll \ell_{\rm max}$ so that the source is
localized enough that the maximum stopping distance may in fact
be determined from the system's response.
As discussed in refs.\ \cite{adsjet,adsjet2}, the effect of a source
like fig.\ \ref{fig:source}b is to generate jets with different
virtualities $-q^2$ on an event-by-event basis.
The typical virtuality of points in the shaded region of
fig.\ \ref{fig:source}b is of order $E/L$, but some have much smaller
virtualities.  The result is a distribution of jet stopping distances
from event to event.  This distribution, as monitored by
where the jet deposits thermalized charge into the system, has
the qualitative form shown in fig.\ \ref{fig:stop} \cite{adsjet}.
The important point for our present purpose is that the distribution
falls exponentially for distances large compared to the maximum
stopping distance scale $\ell_{\rm max} \sim E^{1/3} T^{-4/3}$.
By analyzing quasi-normal modes in the gravity dual, it is
possible to calculate the scale that characterizes the rate of
exponential fall-off.  That is, it is possible to calculate
what we will call ``$\ell_{\rm tail}$'' defined by (\ref{eq:falloff0}),
\begin {equation}
   \mbox{deposition}(x^3)
   \sim
   \mbox{prefactor} \times e^{-x^\three/\ell_{\rm tail}}
   \qquad \mbox{for} \qquad
   x^\three \gg \ell_{\rm max} .
\label {eq:falloff}
\end {equation}
The precise quantitative result for this tail scale $\ell_{\rm tail}$
depends on the choice of
source operator $O(x)$ used in the
source (\ref{eq:source}) to generate the jet.
As an example, for $\lambda{=}\infty$ and for $O(x)$ taken to be a
transverse-polarized R-current, the result for the
deposition tail is \cite{adsjet}%
\footnote{
   In the notation of ref.\ \cite{adsjet}, the $0.539$ in
  (\ref{eq:ltailj}) above is $1/2c_1$.
}
\begin {equation}
   \ell_{\rm tail}(j^\perp) \simeq \frac{0.539 \, E^{1/3}}{(2 \pi T)^{4/3}} \,.
\label {eq:ltailj}
\end {equation}

The tail scale $\ell_{\rm tail}$ is not a perfect stand-in for the
maximum stopping distance $\ell_{\rm max}$.  Consider the case of
scalar source operators with large conformal dimension
$\Delta \gg 1$ but $\Delta$ still parametrically small compared
to powers of $E$ or $\lambda$.
In ref.\ \cite{adsjet2}, it was shown that the corresponding
$\ell_{\rm tail}$ is%
\footnote{
  Specifically, see Appendix D of ref.\ \cite{adsjet2} and
  eqs.\ (D2) and (D20) in particular.
}
\begin {equation}
   \ell_{\rm tail}(\Delta \gg 1) \simeq
   \frac{8 E^{1/3}}{3^{3/2} \Delta^{4/3}}
\end {equation}
but that the exponential fall-off characterized by
(\ref{eq:falloff}) does not begin until a distance
that is parametrically larger in $\Delta$:
\begin {equation}
   \ell_{\rm max} \sim \frac{E^{1/3}}{\Delta^{1/3}} \,.
\end {equation}
This shows an example of how $\ell_{\rm tail}$ can under-estimate
the maximum stopping distance scale.
Nonetheless, $\ell_{\rm tail}$ is a specific, well-defined distance
scale that is relatively easy to compute, and so we will
use it as an example for a precise calculation of a $C^4$
correction related to jet stopping.


\subsection {\boldmath$\ell_{\rm tail}$ from quasi-normal modes}

\begin {figure}
\begin {center}
  \includegraphics[scale=0.4]{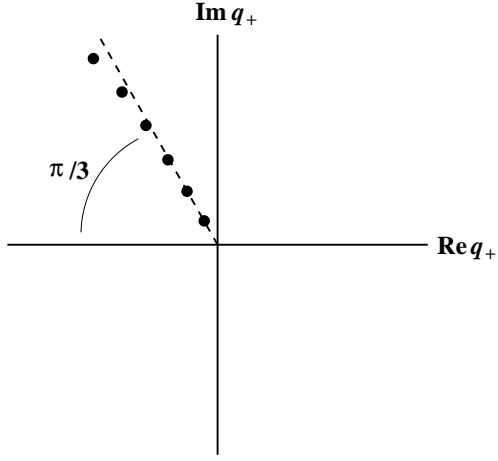}
  \caption{
     \label{fig:poles}
     A qualitative plot of the pole positions of the high-energy,
     retarded boundary-to-bulk propagator in the complex $q_+$ plane.
     Note that the upper-half plane of $q_+ \equiv \frac12(q^\three-q^\zero)$
     corresponds to the lower-half plane of $q^\zero$.
  }
\end {center}
\end {figure}

As discussed in refs.\ \cite{adsjet,adsjet2}, the exponential tail in
fig.\ \ref{fig:stop} is determined by the pole of the
high-energy boundary-to-bulk propagator that is closest to the
real axis in 4-momentum space.%
\footnote{
  For related considerations of poles determining stopping distances,
  in the context of gluon beams created by synchrotron radiation,
  see section VI of ref.\ \cite{CHR}.
}
Specifically, consider the light-cone
components
\begin {equation}
   q^\pm \equiv q^\three \pm q^\zero ,
   \qquad
   q_\pm \equiv \tfrac12 q^\mp = \frac12 (q^\three \mp q^\zero)
\label {eq:lightcone}
\end {equation}
of 4-momentum $q$.  At high energy $E$, the source given by
(\ref{eq:nullkbar}) and
fig.\ \ref{fig:source}b has momenta with
\begin {equation}
   q_- \simeq E .
\end {equation}
The more interesting component of the 4-momentum is therefore $q_+$.
Consider the retarded boundary-to-bulk propagator as a function of $q_+$.
A qualitative sketch of the pole structure in the complex
$q_+$ plane is shown in fig.\ \ref{fig:poles}.%
\footnote{
  When comparing to
  refs.\ \cite{adsjet,adsjet2}, we have canceling differences in
  convention: (i) here we are discussing boundary-to-bulk propagators
  rather than bulk-to-boundary propagators, and (ii) our $q$ here is the
  4-momentum associated with the bulk point rather than the boundary
  point.
}
For large $x^\three$, the exponential tail in
fig.\ \ref{fig:stop} is proportional to \cite{adsjet,adsjet2}
\begin {equation}
   | e^{i q_+^{(0)} x^+} |^2
   \simeq e^{-4 x^\three \Im q_+^{(0)}} ,
\end {equation}
where $q_+^{(0)}$ is the pole nearest the real axis, and deposition
occurs near the light-cone $x^\zero = x^\three$
in position space, so that $x^+ \simeq 2 x^\three$ \cite{adsjet}.
The tail stopping scale $\ell_{\rm tail}$ defined by
(\ref{eq:falloff}) is therefore
\begin {equation}
   \ell_{\rm tail}
   = \frac{1}{4 \Im q_+^{(0)}} .
\label {eq:ltail}
\end {equation}

We now need to find the pole position $q_+^{(0)}$.  The locations of
the poles correspond to quasi-normal modes, which are solutions
that vanish at the boundary and have in-falling boundary
conditions at the horizon, and which are associated with complex
values of $q_+$.


\subsection {Finding quasi-normal mode solutions}

For $\lambda{=}\infty$, the equation of motion for a supergravity
5-dimensional scalar
field in the AdS$_5$-Schwarzschild background is
$(-\nabla^I \nabla_I + m^2) \phi = 0$, which is
\begin {equation}
   -\frac{1}{\sqrt{-g}}
   \partial_\five (\sqrt{-g} \, g^{\five\five} \partial_\five \phi)
   + (q_\mu g^{\mu\nu} q_\nu + m^2) \phi = 0.
\label {eq:eom1}
\end {equation}
The 5-dimensional mass term was unimportant to our previous analysis
based on geometric optics for $\ell_{\rm stop} \ll \ell_{\rm max}$, but it
will be important for the determination of $\ell_{\rm tail}$.
At high energy, the dominant effect of $C^4$-corrections is to modify
the equation of motion (\ref{eq:eom1}) as in (\ref{eq:Qeom2}):%
\footnote{
  \label {foot:eom2}
  To obtain eq.\ (\ref{eq:eom2}), one may pick out the
  most important $C^4$ correction at high energy as outlined
  in section \ref{sec:R4}.  As a check, one may also start from
  a full equation of motion, such as (\ref{eq:hab_eom}) for
  the case $k{=}0$, and pick out the dominant terms for
  (in units where $\zh{=}1$) $|\q| \sim E \gg 1$, $z \sim E^{-1/3}$, and
  $-q^2 \sim E^{2/3}$, which we will see are the relevant scales for
  determining the pole position.
  (This scaling is also described in ref.\ \cite{FestucciaLiu}.)
}
\begin {equation}
   -\frac{1}{\sqrt{-g}}
   \partial_\five (\sqrt{-g} \, g^{\five\five} \partial_\five \phi)
   + (q_\mu g^{\mu\nu} q_\nu + m^2) \phi =
   \frac{\coeff z^{12} |\q|^4}{\zh^8\Rads^2} \, \phi
   ,
\label {eq:eom2}
\end {equation}
with $\coeff$ defined in (\ref{eq:epsilon}).
For the choice (\ref{eq:O}) of source operators in this paper,
which are
dual to the traceless part of $h_{\dot a\dot b}$
and have conformal dimension $\Delta=k+6$,
the 5-dimensional
mass of the dual field in (\ref{eq:eom1}) and (\ref{eq:eom2}) is
given by \cite{KRvN}%
\footnote{
  Our $k=0,1,\cdots$ differs by 2 from the
  $k=2,3,\cdots$ of eq.\ (2.47) of ref.\ \cite{KRvN}, which in turn
  differs by 1 from the $k$ plotted for the $h_{(\alpha\beta)}$ curve
  in fig.\ 2 of ref.\ \cite{KRvN}.
}
\begin {equation}
   (m \Rads)^2
   = \Delta (\Delta - d)
   = (k+6)(k+2) ,
   \qquad
   k=0,1,2,\cdots,
\end {equation}
where $d{=}4$ is the dimension of the boundary theory.

The derivation of the dominant $C^4$ correction also goes through
identically for fluctuations $h_{\one\two}$,
dual to $T^{\one\two}$.
Recall that our convention is that the jet moves in the $x^\three$
direction, and so by $h_{\one\two}$ we mean fluctuations of the form
$h_{\one\two}(x^\zero,x^\three)$ that do not depend on the transverse
coordinates $(x^\one,x^\two)$.  If one writes
$h_{\one\two} = z^{-2} \phi$, then the resulting equation of motion is
identical to (\ref{eq:eom2}) in the high energy limit, but with%
\footnote{
   For comparison, the full $h_{\one\two}$ equation of motion is given
   in (\ref{eq:hxyeq}), and the dominant terms may be picked out
   as in footnote \ref{foot:eom2}.
}
\begin {equation}
   (m \Rads)^2 = 0
   \qquad
   \mbox{for $h_{\one\two}$} .
\end {equation}
On the field theory side of the duality, studying jet stopping with
$T^{\one\two}$ as our source operator can be thought of
as studying the jet that would be created by the decay of a very
high-momentum, slightly off-shell graviton in the 4-dimensional
quark-gluon plasma.

Using the AdS$_5$-Schwarzschild metric (\ref{eq:metric}), the equation of
motion (\ref{eq:eom2}) becomes
\begin {equation}
   -z^3 \partial_z (z^{-3} f \partial_z\phi)
   + \left(
        \frac{q^2}{f} - \frac{z^4 |\q|^2}{\zh^4 f}
        + \frac{(m \Rads)^2}{z^2}
        - \frac{\coeff z^{10} |\q|^4}{\zh^8}
     \right)\phi
   = 0 .
\label {eq:eom3}
\end {equation}
To find the high-energy quasi-normal modes, we will proceed as in
ref.\ \cite{adsjet2}.%
\footnote{
  Specifically appendix C.2 of ref.\ \cite{adsjet2}, but we will
  not make here the $\Delta \gg 1$ approximation used there.
  See ref.\ \cite{FestucciaLiu} for earlier work on high-energy quasi-normal
  modes which used an approximate method to solve the problem.
  Our $\lambda{=}\infty$
  results agree with their approximate results parametrically but
  not in detail.
}
As we will see, the solution for the quasi-normal mode
will be dominated by its behavior at $z \sim E^{-1/3} T^{-2/3} \ll \zh$,
and so we may approximate $f \simeq 1$ in (\ref{eq:eom3}).
Then writing
\begin {equation}
   \phi = z^{3/2}\psi
\end {equation}
gives a Schr\"odinger-like equation
\begin {equation}
   -\tfrac12 \partial_z^2\psi
   + V(z) \psi = -\tfrac12 q^2 \psi
\label {eq:schro1}
\end {equation}
with potential
\begin {equation}
   V(z)
   =
       - \frac{z^4 |\q|^2}{2\zh^4}
       + \frac{(m\Rads)^2+\frac{15}{4}}{2 z^2}
       - \frac{\coeff z^{10} |\q|^4}{2 \zh^8}
   \,.
\end {equation}
We may approximate
$q^2 = 4 q_- q_+ \simeq 4 E q_+$ for our high-energy jets.
Now change variables from $z$ to
\begin {equation}
   Z \equiv e^{-i\pi/6} E^{1/3} \zh^{-2/3} z
\end {equation}
so that (\ref{eq:schro1}) becomes
\begin {equation}
   -\tfrac12 \partial_Z^2\psi
   + {\cal V}(Z) \psi = {\cal E} \psi
\label {eq:schro2}
\end {equation}
with potential
\begin {equation}
   {\cal V}(Z)
   = \frac{{\cal M}^2}{2 Z^2} + \tfrac12 Z^4
     - \tfrac12 \coeff Z^{10} ,
\label {eq:calV}
\end {equation}
\begin {equation}
   {\cal M}^2 \equiv (m\Rads)^2 + \tfrac{15}{4} \,,
\end {equation}
and the ``energy'' ${\cal E}$ in the Schr\"odinger equation
related to $q_+$ by
\begin {equation}
   q_+ = \tfrac12 e^{i 2\pi/3} \zh^{-4/3} E^{-1/3} {\cal E} .
\label {eq:qE}
\end {equation}
We want solutions to the original $\psi(z)$
equation (\ref{eq:schro1}) that,
at large real $z$, only have waves moving away from the boundary
(corresponding to having only in-falling waves at the horizon).
Using the WKB approximation at large $z$, the large-$z$ behavior
for $\coeff = 0$, for example, would be
\begin {equation}
   \psi(z) |_{\coeff{=}0} \sim e^{i z^3 |\q|/3\zh^2} .
\end {equation}
Such solutions
correspond to solutions to the $\psi(Z)$ equation
(\ref{eq:schro2}) that fall exponentially at large real values
of $Z$, e.g.\
\begin {equation}
   \psi(Z) |_{\coeff{=}0} \sim e^{- Z^3/3} .
\end {equation}
Quasi-normal modes are therefore solutions to
(\ref{eq:schro2}) that (i) vanish at $Z{=}0$ and (ii) fall
exponentially for large $Z$.%
\footnote{
   One might worry about the negative sign in the $Z^{10}$
   term in the potential (\ref{eq:calV}), which seemingly implies that
   that ${\cal V}(Z)$ turns over and becomes negative for
   $Z \gg \coeff^{-1/6} \sim \lambda^{1/4}$.
   However, such extremely large values of $Z$ correspond to values of
   $|z|$ larger than the scale $z_{\rm bad}$ of (\ref{eq:zbad})
   where the expansion in higher-derivative corrections breaks down.
   None of these issues will affect the perturbative evaluation of
   the shift in the pole position caused by the $\coeff Z^{10}$
   term.
}
The allowed Schr\"odinger energies ${\cal E}$ in (\ref{eq:schro2}),
which determine the quasi-normal mode values of complex $q_+$ through
(\ref{eq:qE}), are therefore simply the bound state energies associated
with the potential ${\cal V}(Z)$ of
(\ref{eq:calV}) in non-relativistic quantum
mechanics.

Since we've argued in this paper that the $C^4$ correction term is
small and under control for the physics associated with
$\ell_{\rm stop} \sim \ell_{\rm max}$, we should be able to treat
the $\coeff Z^{10}$ term in the potential (\ref{eq:calV}) as
a perturbation.  (We can double check that its effects are small
at the end of the calculation.)  The minimum value of ${\cal E}$,
which determines the location of the $q_+$ pole closest to the
real axis, can then be determined using ordinary perturbation theory
as
\begin {equation}
   {\cal E}_{\rm min} \simeq
   {\cal E}_0
   - \tfrac12 \coeff \langle \psi_0 | Z^{10} | \psi_0 \rangle ,
\label {eq:calE0}
\end {equation}
where ${\cal E}_0$ and $\psi_0(Z)$ are the ground state energy and
wave function for a radial Schr\"odinger equation
associated with a quartic potential $\tfrac12 Z^4$,
with ${\cal M}$ playing the role of angular momentum.
Combining (\ref{eq:calE0}) with (\ref{eq:ltail}) and (\ref{eq:qE}) gives
\begin {equation}
   \ell_{\rm tail} =
   \ell_{\rm tail}^{\lambda{=}\infty}
      \left[ 1 + \coeff \, \frac{\langle Z^{10} \rangle_0}{2 {\cal E}_0}
               + \cdots \right]
\label {eq:ltailgeneral}
\end {equation}
with
\begin {equation}
   \ell_{\rm tail}^{\lambda{=}\infty}
   = \frac{E^{1/3}}{\sqrt3 (\pi T)^{4/3} {\cal E}_0}
   \,.
\label {eq:ltail0}
\end {equation}

Solving for $\psi_0(Z)$ numerically for various $m$'s of interest gives
results for ${\cal E}_0$ and $\langle Z^{10} \rangle_0$ shown in
table \ref{tab:schro}.  Using (\ref{eq:epsilon}) for $\coeff$, the
corresponding formulas for $\ell_{\rm tail}$ are
\begin {subequations}
\label {eq:ltailspecific}
\begin {equation}
   h_{\one\two}: \qquad
   \ell_{\rm tail} =
      \frac{0.3259 \, E^{1/3}}{(2\pi T)^{4/3}}
      \bigl[1 + 47.162 \, \lambda^{-3/2} + O(\lambda^{-5/2})\bigr]
\label{eq:hxytail}
\end {equation}
for source operator $T^{\one\two}$ and
\begin {equation}
   h_{\dot a\dot b}~(k{=}0): \qquad
   \ell_{\rm tail} =
      \frac{0.1704 \, E^{1/3}}{(2\pi T)^{4/3}}
      \bigl[1 + 82.174 \, \lambda^{-3/2} + O(\lambda^{-5/2})\bigr]
\end {equation}
\end {subequations}
for the lowest-dimension
source operator $\tr(\lambda\lambda\bar\lambda\bar\lambda)$
associated with traceless $h_{\dot a\dot b}$.
The first case (\ref{eq:hxytail}) is the result that was previewed in the
introduction (\ref{eq:hxytail0}).

\begin {table}
\begin {tabular}{|c|c|c|c|c|c|c|}
\hline
  source operator & $\Delta$ & R charge rep. & SUGRA field & $(m\Rads)^2$
     & ${\cal E}_0$ & $\langle Z^{10} \rangle_0$ \\
\hline
  $T^{\one\two}$ & 4 & trivial & $h_{\one\two}$ & 0
     & 4.4640 & 58.382 \\
  $\tr(\lambda\lambda\bar\lambda\bar\lambda)$ & 6 & \underline{84}
     & traceless $h_{\dot a \dot b}$ & 12
     & 8.5388 & 194.58 \\
\hline
\end {tabular}
\caption
    {
    \label {tab:schro}
    Numerical results for ${\cal E}_0$ and $\langle Z^{10} \rangle_0$
    for two examples of source operators.  Also listed are the
    conformal dimension $\Delta$ and R-charge representation of the
    source operator, and the corresponding mass-squared of the
    supergravity (SUGRA) field.
    }
\end {table}

We note that in both cases, the results for
$\ell_{\rm tail}^{\lambda{=}\infty}$
are quantitatively fairly well approximated by the
large-${\cal M}$ expression
derived in ref.\ \cite{adsjet2}:%
\footnote{
  Specifically, this comes from the $n{=}0$ case of
  eq.\ (D19) of ref.\ \cite{adsjet2}, which gives what we
  call $2^{-4/3} {\cal E}_0$ in the present paper.  Then use
  (\ref{eq:ltail0}) in the present paper.  The factor of
  $2^{-4/3}$ arises because the definitions of $Z$ in this
  paper and ref.\ \cite{adsjet2} (and so the precise normalization
  of the quartic potential and the Schr\"odinger ``energy'')
  differ by a factor of
  $\zh^{-2/3}$, which is a factor of $2^{2/3}$ in the units
  $2\pi T{=}1$ used in ref.\ \cite{adsjet2}.
}
\begin {equation}
   \mbox{${\cal M}$ large} : \qquad
   \ell_{\rm stop}^{\lambda{=}\infty} \simeq
      \frac{E^{1/3}}{\sqrt3 \, (2\pi T)^{4/3}}
      \left[ \tfrac38 {\cal M}^{4/3}
              + \sqrt{\tfrac38} \, {\cal M}^{1/3} \right]^{-1} .
\end {equation}
However, we find that using the same approximation to calculate
the $\lambda^{-3/2}$ correction term is a very poor approximation,
basically because the large power in $\langle Z^{10} \rangle_0$
is
extremely sensitive to the precise dependence of the wave function on
$Z$.


\section {Conclusion}
\label {sec:conclusion}

At the beginning of this paper, we showed in fig.\ \ref{fig:corrections}
the pattern of the parametric sizes of effects of higher-derivative
supergravity
corrections on the stopping distance.  To demonstrate this pattern, we
looked case by case at the different ways that higher-derivative terms
might appear, and at various possible ways the indices in those terms
might be contracted.
We are suspicious that there must be a quicker, more
elegant way to approach the power counting that would not require
descending to that level of detail.
It would be nice to figure out what the more elegant argument is,
if there is one.

Also, our explicit results (\ref{eq:ltailgeneral}) and (\ref{eq:ltailspecific})
for the leading correction to $\ell_{\rm tail}$ fail to shed light on
a mystery concerning the $\lambda$-dependence of stopping distances:
How does the $E^{1/3}$ scaling of the maximum stopping distance
at $\lambda{=}\infty$
transition to $E^{1/2}$ at small $\lambda$?  One might naively guess
the scaling to be of the form%
\footnote{
   We ignore here logarithmic energy dependence in the prefactor of
   the exponential.  The small-$\lambda$ scaling
   is really $(E/\ln E)^{1/2}$, which is equivalent to including
   a log-of-log
   energy dependence in the exponent $f$:
   $(E/\ln E)^{1/2} = \exp[\tfrac12 - \tfrac12 \ln \ln E]$.
}
\begin {equation}
   \ell_{\rm max} \propto E^{f(\lambda)}
\label {eq:lmaxexponent}
\end {equation}
for some function $f(\lambda)$
with $f(0)=\tfrac12$ and $f(\infty)=\tfrac13$.
One might further hope that $f(\lambda)$ has relatively
simple expansions
around $\lambda{=}0$ and $\lambda{=}\infty$.
For example, perhaps
\begin {equation}
   f(\lambda) = \tfrac12 + \# \lambda + \# \lambda^2 + \cdots
\end {equation}
and
\begin {equation}
   f(\lambda) = \tfrac13 + \# \lambda^{-3/2} + \# \lambda^{-5/2} + \cdots ,
\label {eq:exponent}
\end {equation}
where the $\#$ signs represent numerical coefficients.
The details of the expansion don't matter---one can imagine there could
be different powers of $\lambda$ than shown above, or factors of
$\ln\lambda$ in the expansion, and so forth.
But take (\ref{eq:exponent}) as an example.  Then, expanding
(\ref{eq:lmaxexponent}) around $\lambda{=}\infty$,
\begin {equation}
   \ell_{\rm max} \propto
   E^{1/3} \bigl[ 1 + \# \lambda^{-3/2} \ln E + O(\lambda^{-5/2}) \bigr] .
\end {equation}
We might therefore expect if we compute something  that is related to
the maximum stopping distance, like
$\ell_{\rm tail}$, the corrections in powers of $1/\lambda$
should also come with powers
of $\ln E$, as above.  But there is no sign of a $\ln E$ factor
in (\ref{eq:ltailspecific}).

Perhaps the exponent $f(\infty)=\tfrac13$ does not receive
corrections until a yet-higher power in $\lambda$, but we are unsure
how a $\ln E$ could arise in yet-higher-order calculations of the
shift of the quasi-normal mode pole.  Or perhaps the exponent
does not have an expansion in powers of $1/\lambda$ but
instead behaves like $\tfrac13 + \# e^{-\#\lambda}$ for large
$\lambda$.  Or perhaps the tail scale $\ell_{\rm tail}$ is a misleading
stand-in for $\ell_{\rm max}$, as is known to happen in the case
of $\Delta \gg 1$ \cite{adsjet2}.
Whatever the resolution, given the absence of
a $\ln E$ in our result for $\ell_{\rm tail}$, the question of
how $E^{1/3}$ begins to make its way towards $E^{1/2}$
(and vice versa) remains an open question.


\begin{acknowledgments}

We thank Edmond Iancu for discussions about jet stopping results at
$\lambda{=}\infty$, and we thank
Simon Caron-Huot for discussions about higher-order
corrections at zero temperature.
This work was supported, in part, by the U.S. Department
of Energy under Grant No.~DE-FG02-97ER41027.

\end{acknowledgments}


\appendix


\section{Full \boldmath$C^4$-corrected dispersion relations}
\label {app:dispersion}

In this appendix, we will give examples of the linearized
equation of motion of the metric, to order $O({\alpha'})^4$.
This includes the ${\alpha'}^3$ correction of the
background geometry, as well as the effect of the ${\rm(Weyl)}^4$ term
in the supergravity action.

In this paper, we have generally considered fluctuations
$h_{\dot a \dot b}$ that live on the $S^5$.  There are a variety
of Kaluza-Klein states of this form, but here we will consider those
with the lowest mass, which transform under the
$(2,0,2) = \underline{84}$ representation of SU(4) \cite{KRvN}.
But we will also give similar results for fluctuations $h_{\one\two}$
of the AdS$_5$-Schwarzschild metric.
For simplicity, we set $\zh=1$ and $\Rads=1$ throughout this
appendix.

We start with $h_{\dot a \dot b}$.
In SO(6) language, the Young tableau of the $\underline{84}$
has the form of a box, with two rows and two columns.
The corresponding spherical harmonic on $S^5$ is given by
\begin{eqnarray}
  h_{\dot a\dot b}&=&
  \phi^{\bf ABCD}(x)\bigg[
       V^{\bf A B}_{((\dot a}V^{\bf CD}_{\dot b))}
\nonumber\\
  &&  -\frac{1}{N-2}\bigg(
          \delta^{\bf A C} V^{\bf EB}_{((\dot a} V^{\bf ED}_{\dot b))}
        + \delta^{\bf BD} V^{\bf AE}_{((\dot a} V^{\bf CE}_{\dot b))}
        - \delta^{\bf BC} V^{\bf EA}_{((\dot a} V^{\bf ED}_{\dot b))}
        - \delta^{\bf AD} V^{\bf EB}_{((\dot a} V^{\bf EC}_{\dot b))}
      \bigg)
\nonumber\\
  &&+\frac{1}{(N-1)(N-2)}
      \bigg(\delta^{\bf AC}\delta^{\bf BD}
           -\delta^{\bf BC}\delta^{\bf AD}\bigg)V^{\bf EF}_{((\dot a}
       V^{\bf EF}_{\dot b))}
  \bigg],
\end{eqnarray}
where the double round bracket notation stands for symmetrization and
removal of a trace, $x$ denotes the non-compact coordinates, and $N=6$
for $SO(6)$, which is the isometry group of $S^5$. We have also defined
the embedding scalar functions
\begin{eqnarray}
  &&Y^{\bf A} Y^{\bf A}=1,\qquad {\bf A}=1,2,\dots 6,
\nonumber\\
  &&Y^{\bf1}+iY^{\bf 2}=\sin(\theta_1)\cos(\theta_2)e^{i\phi_1},
    \;\; Y^{\bf3}+iY^{\bf 4}=\sin(\theta_1)\sin(\theta_2)e^{i\phi_2},
    \;\; Y^{\bf5}+iY^{\bf 6}=\cos(\theta_1)e^{i\phi_3}
    ,
\nonumber\\
\end{eqnarray}
and written the Killing vectors in terms of them,
\begin{eqnarray}
  V^{\bf AB}_{\dot c}=
  V^{\bf [AB]}_{\dot c}=Y^{[\bf A}\partial_{\dot c}Y^{\bf B]}.
\end{eqnarray}
The reason for choosing such a fluctuation is twofold. Firstly,
$h_{\dot a\dot b}$ originates in a traceless metric fluctuation on $S^5$, and it
does not mix to quadratic order with fluctuations of other
fields. Secondly, in case one is interested in setting up a conserved
charge measurement as in \cite{adsjet}, $h_{\dot a \dot b}$, with its
nontrivial spherical harmonic on the sphere, can be used as source for a
jet carrying R-charge.

The equation of motion of one of the $\underline{84}$ scalars
is
\begin{eqnarray}
  &&0=
  \frac{f(z)}{2z^3} \frac{d^2 \phi(z)}{dz^2}
  -\frac{1}{2z^4}(3+z^4)\frac{d \phi(z)}{dz}
  -\frac{1}{2z^3}\phi(z) \bigg(
      \frac{12}{z^2}+\q^2-\frac{E^2}{f(z)}
    \bigg)
\nonumber\\ &&\;\;
  +\gamma\bigg[
    \frac{57}{8}\frac{d^4\phi(z)}{dz^4} f^2(z) z^7
   -\frac{57}{4} \frac{d^3\phi(z)}{dz^3} f(z)(15 z^4-7) z^6
\nonumber\\ &&\quad
  +\frac{z}{16} \frac{d^2 \phi(z)}{dz^2}
    \bigg(
      3(9063 z^{12} -9668 z^8 +2021 z^4 -200)+220z^6 f(z) \q^2+228 z^6 E^2
    \bigg)
\nonumber\\ &&\quad
  +\frac {1}{16}\frac{d\phi(z)}{dz}\bigg(
      3(14079 z^{12}-11772 z^8 +2125 z^4 -200)
      -220 z^6(11 z^4-7)\q^2
      +1596 z^6 E^2
    \bigg)
\nonumber\\ &&\quad
  -\frac{z \phi(z)}{16 f^2(z)}\bigg(
     -36 (30z^8-59 z^4+269) z^2 f^2(z)
     +(2025 z^4-1799)z^4 f^2(z)\q^2
\nonumber\\ &&\quad\hspace{10em}
  -(1023 z^{12}-3504 z^8 +3705 z^4 +600)E^2
\nonumber\\ &&\quad\hspace{10em}
  -220 z^6f(z)\q^2 E^2
     - 50 z^6f^2(z) (\q^2)^2 -114 z^6 E^4
  \bigg)\bigg],
\label{eq:hab_eom}
\end{eqnarray}
where $\gamma$ is as defined in \cite{BLS}:
\begin{equation}
  \gamma=\tfrac18\zeta(3){\alpha'}^3 .
\end{equation}
This equation may be solved perturbatively in
$\gamma$: $\phi(z)=\phi_0+\gamma\phi_1 +\cdots$.
After substituting the solution of the zeroth
order in $\gamma$ equation of motion, $\phi_0$, (\ref{eq:hab_eom}) becomes
\begin{eqnarray}
  &&0= \frac{f(z)}{2z^3} \frac{d^2 \phi_1(z)}{dz^2}
   -\frac{1}{2z^4}(3+z^4)\frac{d \phi_1(z)}{dz}
   -\frac{1}{2z^3}\phi_1(z) \bigg(
      \frac{12}{z^2}+\q^2-\frac{E^2}{f(z)}
   \bigg)
\nonumber\\
  &&\;\; +\frac{f(z)}{4}\frac{d\phi_0(z)}{dz} \bigg(
       3(-1767z^8+4326z^4-200) +1120 z^6 \q^2
    \bigg)
\nonumber\\
  &&\;\; +\frac{\phi_0(z)}{2z}\bigg(9(15z^{12}-1540 z^8+1596 z^4-100)
\nonumber\\
  &&\;\; -z^2 (1899 z^8-2221 z^4+75)\q^2
     +6 z^2(101 z^8-131 z^4 +25)\frac{E^2}{f(z)}
     +48 z^8 (\q^2)^2\bigg).
\end{eqnarray}

Now turn to the linearized equation of motion for $h_{\one\two}$.
If one defines
$\phi(x) = h_{\one\two} z^2$, then similar considerations yield
\begin{eqnarray}
  &&0=
  \frac{2f(z)}{z}\frac{d^2\phi(z)}{dz^2}
  -\frac{2}{z^2}(3+z^4)\frac{d\phi}{dz}
  -\frac{2}{z}\phi\bigg(\q^2 - \frac{E^2}{f(z)}\bigg)
\nonumber\\ &&\;\;
  +\gamma\bigg[\frac{37}{2} \frac{d^4\phi(z)}{dz^4} f^2(z) z^9
  - 37\frac{d^3\phi(z)}{dz^3} f(z)z^8 (15 z^4-7)
\nonumber\\ &&\quad
  +\frac{z^3}{4}\frac{d^2\phi(z)}{d^2z}\bigg(
      15445 z^{12}-15340 z^8+2863 z^4-600
      +236 f(z)z^6 \q^2+148 z^6 E^2
    \bigg)
\nonumber\\ &&\quad
  +\frac{z^2}4\frac{d\phi(z)}{dz}\bigg(
      -1235z^{12}+8460 z^8-1225 z^4-600-
      236z^6(11z^4-7)\q^2+1036z^6 E^2
    \bigg)
\nonumber\\ &&\quad
  +\frac{z^3\phi(z)}{4 }\bigg(
      -200z^2(9z^8-2z^4-2)-z^4(465 z^4-719)\q^2
\nonumber\\ &&\quad\hspace{7em}
      +(-2665z^{12}+3968 z^8-719 z^4+600)\frac{E^2}{f^2(z)}
\nonumber\\ &&\quad\hspace{7em}
  +236 z^6\q^2\frac{E^2}{f(z)}
     +74 z^6  (\q^2)^2
     +74 z^6\frac{E^4}{f^2(z)}
   \bigg)\bigg]
\label {eq:hxyeq}
\end{eqnarray}
(This equation extends that of ref.\ \cite{BLS}
to the case of non-zero $\q$.)%
\footnote{
   In particular, setting $\q = 0$ in our eq.\ (\ref{eq:hxyeq})
   reduces it to eq.\ (3.25) of ref.\ \cite{BLS}, after fixing an
   obvious typo in the latter.
}
Solving for $\phi=\phi_0+\gamma\phi_1+\dots$ order-by-order in $\gamma$
yields the following equation of motion for $\phi_1$:
\begin{eqnarray}
  &&0=\frac{2f(z)}{z}\frac{d^2\phi_1(z)}{dz^2}
    -\frac{2}{z^2}(3+z^4)\frac{d\phi_1}{dz}-
    \frac{2}{z}\phi_1\bigg(\q^2 - \frac{E^2}{f(z)}\bigg)
\nonumber\\ &&\;\;
  +\frac{d\phi_0(z)}{dz}f(z) z^2(3171 z^8+2306 z^4-600 + 960 z^6\q^2)
\nonumber\\ &&\;\;
  +\phi_0(z)\bigg(
      -50(9z^8-2z^4-2)z^5+12(89z^8-119 z^4+25)\frac{E^2}{f(z)}
\nonumber\\ &&\;\;\hspace{5em}
  -2(851z^8-789 z^4+75)z^3\q^2 + 96 z^9 (\q^2)^2
     \bigg).
\end{eqnarray}


\section{Why (\ref{eq:importanceC4}) cannot precisely
         determine \boldmath$\Delta \ell_{\rm stop}$}
\label {app:C4}

Consider the safe region
$\ell_{\rm stop} \gg \lambda^{-1/6} \ell_{\rm max}$
of fig.\ \ref{fig:corrections}, where the effects of
higher-dimensional supergravity interactions should be suppressed.  The $R^4$
corrections then dominate the corrections at $z \sim z_\star$.
We might then be tempted to use the explicit
form (\ref{eq:C4}) of the $R^4$ correction, combined with the
particle-based formula (\ref{eq:stopq5}) for the stopping distance, to
explicitly calculate the first correction to the $\lambda{=}\infty$
result (\ref{eq:stop}) for the stopping distance.
In this appendix, we will discuss why that does not work.

We will start with (\ref{eq:stopC4}),
\begin {equation}
   \ell_{\rm stop} \simeq
   \int_0^{\zh} dz \>
   \frac{
     |\q| \left[ 1 - \frac{2 \coeff z^{10}}{\zh^8} |\q|^2 \right]
   }{
     \sqrt{
          - q^2 + \frac{z^4}{\zh^4} |\q|^2
          + \frac{\coeff z^{10}}{\zh^8} |\q|^4 f
     }
   } ,
\label {eq:stopC4app}
\end {equation}


\subsection {Numerator correction}

We earlier promised a discussion of why the potentially sign-changing
behavior of the numerator correction in
the $R^4$-corrected formula (\ref{eq:stopC4app}) for the stopping distance
could be ignored.  The disturbing features of this correction arise
in the $z$ range given by (\ref{eq:C4problem}),
\begin {equation}
  z \gg z_{\rm disturbing}
  \sim
  \left( \frac{\lambda^{3/4} T}{E} \right)^{1/5} \zh .
\label {eq:zdisturbing}
\end {equation}
First of all, notice that this difficulty only arises at all if
$z_{\rm disturbing} < \zh$, which requires
\begin {equation}
  E \gg \lambda^{3/4} T .
\label {eq:Edisturb}
\end {equation}
Now compare (\ref{eq:zdisturbing}) and (\ref{eq:zbad}):
\begin {equation}
  z_{\rm disturbing} \sim
  \left( \frac{E}{\lambda^{1/8} T} \right)^{2/15} z_{\rm bad} .
\end {equation}
The inequality (\ref{eq:Edisturb}) then gives
\begin {equation}
  z_{\rm disturbing} \gg
  z_{\rm bad} ,
\end {equation}
and so the numerator correction cannot be believed in the range of $z$
for which it becomes disturbing.


\subsection {Denominator correction}

Dropping the numerator correction from (\ref{eq:stopC4app}) leaves
\begin {equation}
   \ell_{\rm stop} \simeq
   \int_0^{\zh} dz \>
   \frac{
     |\q|
   }{
     \sqrt{
          - q^2 + \frac{z^4}{\zh^4} |\q|^2
          + \frac{\coeff z^{10}}{\zh^8} |\q|^4 f
     }
   } ,
\label {eq:stopC4b}
\end {equation}
For simplicity, in what follows we will just analyze the case
$E \gg \lambda^{3/4} T$.

In the integrand, look at the expression under the square root in the
denominator.  The relative importance of the $\coeff z^{10}$ term grows
with increasing $z$.  At what $z$ scale does it start to dominate?
We know that, when the $R^4$ corrections are small,
the $\coeff z^{10}$ term is a small correction at $z \sim z_\star$.
But what about at larger $z$?
For $z \gg z_\star$, the $z^4$ term under the square root dominates
over the $-q^2$ term, so we should compare the $\coeff z^{10}$
term to the $z^4$ term.  These are the same size at a scale
$z_{\star\star} \gg z_\star$ given by
\begin {equation}
   z_{\star\star}
   \sim \frac{\zh^{2/3}}{\coeff^{1/6} |\q|^{1/3}}
   \sim \frac{\lambda^{1/4}}{E^{1/3} T^{2/3}}
   \sim \frac{\lambda^{1/4}}{l_{\rm max} T^2} \,,
\label {eq:zsstar}
\end {equation}
assuming that $z_{\star\star} \ll \zh$ so that $f \simeq 1$.
But $z_{\star\star} \ll \zh$ follows from (\ref{eq:zsstar}) and
our consideration of $E \gg \lambda^{3/4} T$.

Now calculate the correction $\Delta\ell$ to the stopping distance
by subtracting the $\lambda{=}\infty$ result (\ref{eq:stop1})
from (\ref{eq:stopC4b}),
\begin {equation}
   \Delta \ell \equiv
   \ell_{\rm stop} - \ell_{\rm stop}^{\lambda{=}\infty}
   \simeq
   \int_0^{\zh} dz
   \left[
     \frac{|\q|}
       { \sqrt{ - q^2 + \frac{z^4}{\zh^4} |\q|^2
                + \frac{\coeff z^{10}}{\zh^8} |\q|^4 f }
       }
     -
     \frac{|\q|}
       { \sqrt{ - q^2 + \frac{z^4}{\zh^4} |\q|^2 } }
   \right] .
\label {eq:Deltal}
\end {equation}
This integral is dominated by $z \sim z_{\star\star}$.
So, to explicitly
calculate $\Delta \ell$ will require trusting the integrand
at $z \sim z_{\star\star}$ given by (\ref{eq:zsstar}).
Compare this to the $z$ scale (\ref{eq:zbad}) where the
expansion in supergravity corrections breaks down:
\begin {equation}
   z_{\star\star} \sim \lambda^{1/12} z_{\rm bad} \gg z_{\rm bad} .
\end {equation}
So we cannot trust (\ref{eq:Deltal}) in the range of $z$ where we
want to use it to get an explicit result for $\Delta\ell$.

Note that, in order to use a particle-based formula such as
(\ref{eq:Deltal}), one should check
that the geometric optics approximation is actually valid
at $z \sim z_{\star\star}$.
Arguments similar to those of ref.\ \cite{adsjet2} show that,
for the $\lambda{=}\infty$ calculation, the geometric optics
approximation is okay for $z_{\rm WKB} \ll z \ll z_{\rm wave}$
where $z_{\rm WKB} \sim 1/\sqrt{-q^2}$ and
$z_{\rm wave} \sim -q^2/ET^2$.
But discussing the resulting constraints is moot since we already
have other reasons
not to believe (\ref{eq:Deltal}) at $z \sim z_{\star\star}$.


\section {Other higher-derivative terms}
\label {app:details}

\subsection{\boldmath$\alpha' Q^\mu D_\mu$ contributions to
                 \boldmath$\alpha' D^2$}

In section \ref{sec:D2cost}, we focused on the
$\alpha' Q^\five D_\five$ piece of $\alpha' D^2$ when studying the importance of
$D^{2n} C^4$.  We also recycled our conclusion from that analysis
when later considering
applying extra powers of derivatives to $D^{2k} C^{4+k}$.
The dominant terms involved $\alpha' Q^I D_I$ where the $D_I$ hits
a background Weyl tensor.  We motivated focusing on $Q^\five D_\five$
by noting that the background Weyl tensor depends only of the
$x^\five$ coordinate.  If the $D$'s were ordinary derivatives instead
of covariant derivatives, that would be the end of the story.  However,
the other components $D_\mu$ of the covariant derivative do not vanish
when applied to the background Weyl tensor.  In fact, they are parametrically
of order $1/z$, just like $D_\five$.  As a result, for example,
\begin {equation}
   \alpha' Q^\three D_\three
   = \alpha' Q_\three g^{\three\three} D_\three
   \sim \alpha' \times E \times \frac{z^2}{\Rads^2} \times \frac{1}{z}
   \sim \frac{E z}{\lambda^{1/2}}
\label {eq:QD3}
\end {equation}
is actually parametrically {\it larger}\/ than the derivative
\begin {equation}
   \alpha' Q^\five D_\five
   \sim \frac{q_\five z}{\lambda^{1/2}}
\label {eq:QD5}
\end {equation}
considered in the main text (\ref{eq:D2cost1}).

So why doesn't this lead to much larger results for the importance
of $D^{2n} R^4$ and other operators than shown in fig.\ \ref{fig:corrections}?
Our answer requires thinking about how the indices of the
background Weyl tensor $C_{JKLM}$ hit by $\alpha' Q^I D_I$ contract with
everything else.

Because $C_{IJKL}$ depends only on $x^\five$,
a non-zero value for $Q^\mu D_\mu C_{IJKL}$ arises only from the
terms of $D$ involving the Christoffel symbols:
\begin {equation}
   Q^\mu D_\mu C_{IJKL} =
   -Q^\mu {\Gamma^{\bar I}}_{I\mu} C_{\bar IJKL}
   -Q^\mu {\Gamma^{\bar J}}_{J\mu} C_{I\bar JKL}
   -\cdots .
\end {equation}
Now write
\begin {equation}
   \Gamma = \Gamma^{\rm(AdS)} + \Delta\Gamma ,
\label {eq:Gamma}
\end {equation}
where $\Gamma^{\rm(AdS)}$ is the zero-temperature, purely AdS expression
for the connection $\Gamma$.  The difference between AdS and
AdS$_5$-Schwarzschild is the difference between taking $f=1$ and
$f=1-(z/\zh)^4$ in the metric (\ref{eq:metric}).
As a result, the $\Delta\Gamma$ piece of (\ref{eq:Gamma})
is suppressed compared to the $\Gamma^{\rm(AdS)}$ piece by
order $(z/\zh)^4$.  For studying the dominant corrections
at $z\sim z_\star \ll \zh$, we should therefore focus on
$\Gamma^{\rm(AdS)}$.  In particular,
\begin {equation}
   \alpha' Q^\mu {\Delta\Gamma^{\bar I}}_{I\mu}
   \sim \alpha' E \times \frac{z^2}{\Rads^2}
      \times \frac{1}{z}\left(\frac{z}{\zh}\right)^4
   \sim \frac{E z^5}{\lambda^{1/2} \zh^4}
\end {equation}
is always less important at $z \sim z_\star$ than the
$\alpha' Q^\five D_\five$ term (\ref{eq:QD5}) that we considered in
the main text.

So now focus on $\Gamma^{\rm AdS}$:
\begin {equation}
   Q^\mu D_\mu C_{IJKL} \simeq
   - Q^\mu ({\Gamma^{\bar I}}_{I\mu})^{\rm(AdS)} C_{\bar IJKL}
   - Q^\mu ({\Gamma^{\bar J}}_{J\mu})^{\rm(AdS)} C_{I\bar JKL}
   - \cdots .
\end {equation}
Because zero-temperature, purely AdS space has 4-dimensional
Lorentz invariance, the $\mu$ index on $Q^\mu$ above must pass
through to contract with something else.  For example,
\begin {align}
   Q^\mu ({\Gamma^{\bar I}}_{I\mu})^{\rm(AdS)} & C_{\bar IJKL}
      \times (\mbox{other stuff})^{IJKL}
\nonumber\\ &
   \simeq
   \frac{1}{z} \, Q^\mu C_{\mu JKL} \times (\mbox{other stuff})^{\five JKL}
   - \frac{1}{z} \,
       {C^\five}_{JKL} Q_\mu \times (\mbox{other stuff})^{\mu JKL}
\label {eq:example1}
\end {align}
and
\begin {align}
   Q^\mu ({\Gamma^{\bar J}}_{J\mu})^{\rm(AdS)} & C_{I\bar JKL}
      \times (\mbox{other stuff})^{IJKL}
\nonumber\\ &
   \simeq
   \frac{1}{z} \, Q^\mu C_{I\mu KL} \times (\mbox{other stuff})^{I\five KL}
   - \frac{1}{z} \,
       {{C_I}^\five}_{KL} Q_\mu \times (\mbox{other stuff})^{I\mu KL}
\label {eq:example2}
\end {align}

But now recall that our dominant terms already had every $C$ contracted
with two $Q$'s.  So the ``other stuff'' above had the form
\begin {equation}
   (\mbox{other stuff})^{IJKL} \sim Q^I Q^K (\mbox{something})^{JL} ,
\end {equation}
and these terms were dominant because both $Q_I$ and $Q_K$ were
parametrically of order $E$ when contracted with the Weyl tensor
$C_{IJKL}$.  We are currently worried about the possibility that
the $Q_\mu$ factor above is also of order $E$.  Now look at the first
term in (\ref{eq:example1}).  The $Q_\mu Q_I Q_K \sim E^3$ is contracted
in such a way that it instead gives
$Q_\mu Q_\five Q_K \sim q_\five E^2 \ll E^3$, which is not problematical.
The second term in (\ref{eq:example1}) contracts two $Q$'s together
to give a factor of $Q_\mu \eta^{\mu\nu} Q_\nu \sim q^2$ instead of an $E^2$,
and so it also is suppressed.
Now look at the first term in (\ref{eq:example2}).  There we have
$Q^\mu Q^I Q^K C_{I\mu KL}$.  Up to terms which are suppressed by
$q_\five \ll E$, this is the same as $Q^J Q^I Q^K C_{IJKL}$, which
vanishes by the symmetry of the Weyl tensor.  Finally, look at the
second term of (\ref{eq:example2}), which involves
\begin {equation}
   Q_\mu (\mbox{something})^{\mu L} .
\end {equation}
For the dominant terms analyzed in the main text of this paper,
the ``something'' is made
up of factors of $Q$ and $QQC$.  If $(\mbox{something})^{\mu L}$
gives a factor of $Q^\mu$, then two of our $Q$'s that were supposed
to be giving factors of $E$ will instead give a factor of $-q^2 \ll E$.
If $(\mbox{something})^{\mu L}$ gives a factor of
$Q_N Q_P C^{N \mu P \bul}$, then we'll get a suppression as before because
of the symmetry of $C$.

The upshot is that the worrisome $\alpha' Q^\mu D_\mu$ pieces of
$\alpha' D^2$ seem large in isolation but do
not give large contributions when combined with the rest of
the expression that they are a part of.  The $Q^\mu$ has to contract with
something else in the expression.  The only other things there at
leading order in energy are other factors of $Q$ and $QQC$.  In either
case, this contraction leads to additional suppressions so that
$\alpha' Q^\mu D_\mu$ does not give anything larger than what we
already considered in the main text.  The same reasoning applies
to higher powers $(\alpha' Q^\mu D_\mu)^k$.

The argument in this sub-section is particularly lacking in elegance.
As mentioned in
the conclusion, it would be nice to devise a
more elegant argument.


\subsection{\boldmath$C^2$ and \boldmath$C^3$}
\label{sec:C2C3}

There are no $C^2$ and $C^3$ corrections to supergravity from
tree-level Type II string
theory, but it will make a later argument smoother if we show that
such corrections would not change things, even were they present.
An $\alpha' C^2$ correction would generate a quadratic term for
$\phi$ of the form
\begin {equation}
  \alpha' (\nabla\nabla \phi) (\nabla\nabla\phi) ,
\end {equation}
which in turn would modify the dispersion relation to something of
the form
\begin {equation}
   Q^I Q_I =
   \alpha' g^{\bul\bul} g^{\bul\bul} Q_\bul Q_\bul Q_\bul Q_\bul + \cdots \,.
\end {equation}
That is,
\begin {equation}
   Q^I Q_I =
   \alpha' (Q^I Q_I)^2 + \cdots ,
\label{eq:C2disp}
\end {equation}
where the ellipses indicates other corrections like the effect of $C^4$.
The $C^2$ correction displayed explicitly in (\ref{eq:C2disp})
will turn out to be relatively innocuous because the $Q$'s had to
be contracted together and $Q^I Q_I = 0$ for $\lambda{=}\infty$.
Let's compare the $\alpha' (Q^I Q_I)^2$ term on the right-hand side
of (\ref{eq:C2disp}) to
the corresponding term from $C^4$ shown in (\ref{eq:Qeom2}).  Using
(\ref{eq:QQ}), the ratio of the two is
\begin {equation}
   \frac{\alpha' (Q^I Q_I)^2}{\alpha'^3 z^{12}E^4/(\zh \Rads)^8}
   \biggr|_{z\sim z_\star}
   \sim
   \frac{\alpha'^4 z_\star^{12}E^4}{(\zh \Rads)^8}
   \sim
   \frac{(-q^2)^3}{\lambda^2 E^2 T^4}
   \sim
   \left( \frac{\lambda^{-1/6} \ell_{\rm max}}{\ell_{\rm stop}} \right)^{12}
   .
\label {eq:C2fac}
\end {equation}
That is, $C^2$ is much less important than $C^4$ for
$\ell_{\rm stop} \gg \lambda^{-1/6} \ell_{\rm max}$.
If we multiply (\ref{eq:C2fac}) by the importance
(\ref{eq:importanceC4}) of $C^4$, we can summarize as
\begin {equation}
  \mbox{Importance($C^2$)}
  \sim
  \lambda^{-1/2}
  \left( \frac{ \lambda^{-1/6} \ell_{\rm max} }{ \ell_{\rm stop} }
       \right)^{18} .
\end {equation}

Now make the same analysis for the effect that an
$\alpha'^2 C^3$ term in the supergravity action would have.
The corresponding quadratic term for $\phi$ would be
\begin {equation}
  \alpha'^2 (\nabla\nabla \phi) (\nabla\nabla\phi) C .
\end {equation}
Let's first follow the naive reasoning that the dominant term will
be the one where the derivatives all give factors of $Q$.
Then the dispersion relation takes the form
\begin {equation}
   Q^I Q_I =
   \alpha'^2
   g^{\bul\bul} g^{\bul\bul}
   g^{\bul\bul} g^{\bul\bul}
   Q_\bul Q_\bul Q_\bul Q_\bul C_{\bul\bul\bul\bul}
   + \cdots \,.
\label {eq:C3disp1}
\end {equation}
Because of the symmetry of the Weyl tensor, two of the $Q$'s must
contract with each other.  So the term shown on the
right-hand side has size
\begin {equation}
   \alpha'^2
     Q^I Q_I g^{\bul\bul} g^{\bul\bul} g^{\bul\bul}
     Q_\bul Q_\bul C_{\bul\bul\bul\bul}
   \sim
   \alpha'^2 Q^I Q_I
     \times \left( \frac{z^2}{\Rads^2} \right)^3
     \times E^2 \times \frac{\Rads^2}{\zh^4} \,.
\end {equation}
This term will also be relatively innocuous because of the
contracted factor of $Q^I Q_I$.  Using (\ref{eq:QQ}),
its importance relative to the corresponding $C^4$ term shown
in (\ref{eq:Qeom2}) is
\begin {equation}
   \frac{\alpha'^2 Q^I Q_I z^6 E^2 / (\zh \Rads)^4}
        {\alpha'^3 z^{12} E^4/(\zh \Rads)^8}
   \biggr|_{z\sim z_\star}
   \sim
   \frac{\alpha'^2 z_\star^6 E^2}{\zh^4 \Rads^4}
   \sim
   \frac{(-q^2)^{3/2}}{\lambda E T^2}
   \sim
   \left( \frac{\lambda^{-1/6} \ell_{\rm max}}{\ell_{\rm stop}} \right)^{6} .
\end {equation}
So $C^3$, if there were such a correction,
would also be much less important than $C^4$ for
$\ell_{\rm stop} \gg \lambda^{-1/6} \ell_{\rm max}$,
and
\begin {equation}
  \mbox{Importance($C^3$)}_{\,{\rm all}~Q{\rm's}}
  \sim
  \lambda^{-1/2}
  \left( \frac{ \lambda^{-1/6} \ell_{\rm max} }{ \ell_{\rm stop} }
       \right)^{12} .
\end {equation}

However, we learned in section \ref{sec:D2cost} that factors $Q^I Q_I$
of contracted $Q$'s are often beaten by contributions where one of
the derivatives acts instead on the background curvature.
In addition to (\ref{eq:C3disp1}), we also have contributions of the
form
\begin {align}
   Q^I Q_I &=
   \alpha'^2
   g^{\bul\bul} g^{\bul\bul}
   g^{\bul\bul} g^{\bul\bul}
   Q_\bul Q_\bul \nabla_\bul \nabla_\bul C_{\bul\bul\bul\bul}
   + \cdots
\nonumber\\
   &\sim
   \alpha'^2
     \times \left( \frac{z^2}{\Rads^2} \right)^4
     \times E^2 \times z^{-2} \times \frac{\Rads^2}{\zh^4}
     + \cdots.
\end {align}
Compared to the corresponding $C^4$ contribution, the term shown on
the right-hand side has relative importance
\begin {equation}
   \frac{\alpha'^2 z^6 E^2 / \zh^4 \Rads^6}
        {\alpha'^3 z^{12} E^4/(\zh \Rads)^8}
   \biggr|_{z\sim z_\star}
   \sim
   \frac{\zh^4 \Rads^2}{\alpha' z_\star^6 E^2}
   \sim
   \frac{\lambda^{1/2} E T^2}{(-q^2)^{3/2}}
   \sim
   \lambda^{-1/2}
   \left( \frac{\lambda^{-1/6} \ell_{\rm max}}{\ell_{\rm stop}} \right)^{-6} .
\end {equation}
Multiplying by the importance (\ref{eq:importanceC4}) of $C^4$ gives
\begin {equation}
  \mbox{Importance($C^3$)}_{\,{\rm two}~Q{\rm's}}
  \sim
  \lambda^{-1/2} ,
\end {equation}
which is always small.


\subsection{Aside on Gauss-Bonnet gravity}
\label {sec:GB}

In the last sub-section, our evaluation of the importance of $C^2$
used the non-zero value (\ref{eq:QQ}) of $Q^I Q_I$ that is induced by the
$C^4$ term.
Though not relevant to our study of ${\cal N}{=}4$ SYM, some readers
may be curious what would happen if instead there were {\it only}\/
$R^2$ corrections and all other higher-derivative corrections were
absent.%
\footnote{
  For a discussion of the effects that Gauss-Bonnet gravity would have
  on other types of ``jet'' stopping calculations, such as drag
  forces on trailing classical
  strings or calculations of $\hat q$, see refs.\ \cite{Fadafan1,Fadafan2,NGT}.
}
If we correspondingly
throw away the ``$\cdots$'' in the $C^2$ dispersion relation
(\ref{eq:C2disp}) (and we'll see in the next sub-section that it doesn't
matter whether it is $C^2$ or $R^2$), then $Q^I Q_I = 0$ remains a
solution to the dispersion relation.  So the geometric-optics based
calculation of stopping distances can in this case be accomplished with the
simple null geodesic formula (\ref{eq:null})---the only difference is
that one should use the $R^2$-corrected background metric instead
of the unperturbed AdS$_5$-Schwarzschild metric.

One might in particular consider the case of the
$R^2$ correction to ${\cal N}{=}2$ $D{=}5$ gauged
supergravity (in contrast to the IIB string theory corrections to
${\cal N}{=}8$ $D{=}5$ gauged supergravity relevant to this paper).
Working perturbatively in the $R^2$ coupling these terms may be put into
a Gauss-Bonnet form by a field redefinition of the metric.
Let $\lambda_{\rm GB}$ parametrize the coefficient of the
Gauss-Bonnet $R^2$ term in the supergravity action.
Then, assuming all yet-higher derivative terms in the
Lagrangian are absent,
the correction to the AdS$_5$-Schwarzschild background is known
to all orders in $\lambda_{\rm GB}$ \cite{Cai,BEMPSS}.
If one carries through the null
geodesic calculation in this modified metric, it turns out that
one finds the same
$\ell_{\rm stop} \propto (-q^2/E^2)^{1/4}/T$ result as in
(\ref{eq:stop}), but the constant of proportionality is a function of
$\lambda_{\rm GB}$.%
\footnote{
   Specifically, for $\lambda_{\rm GB}$ as defined in ref.\
   \cite{BEMPSS} for $D{=}5$, we find (\ref{eq:stop}) is modified by an overall
   factor of
   $[s(1+s)/2]^{1/4}$ where $s \equiv \sqrt{1-4\lambda_{\rm GB}}$.
}


\subsection{\boldmath$D^{2n} R^m$ vs.\ \boldmath$D^{2n} C^m$}
\label {sec:R0}

In the main text, we considered only powers of the Weyl curvature tensor
rather than the Riemann curvature tensor.  Rather than spend time
wondering whether one or the other arises from string interactions
at all powers of curvature,
we can dispense with the difference with a relatively simple argument.
At first sight, using Riemann instead of Weyl might seem to make
a dramatic difference in our previous power counting arguments because
\begin {equation}
   R_{IJKL} \sim \frac{\Rads^2}{z^4}
   \qquad \gg \qquad
   C_{IJKL} \sim \frac{\Rads^2}{\zh^4}
\end {equation}
for the case $z \sim z_\star \ll \zh$ of interest.
But, even if there are such higher-derivative terms in the action with
$R$'s rather than $C$'s, they do not cause a problem.

To see this, separate out the $z^{-4}$ piece of the Riemann tensor,
which is just the {\it zero}-temperature Riemann tensor corresponding
to AdS rather than AdS$_5$-Schwarzschild.  But the AdS Riemann tensor
has a simple form in terms of the AdS metric:
\begin {equation}
   R^{\rm AdS}_{\bul\bul\bul\bul}
   \propto g^{\rm AdS}_{\bul\bul} g^{\rm AdS}_{\bul\bul}
           ~\mbox{(appropriately symmetrized)}
   .
\end {equation}
The idea is to separate this piece from the Ads-Schwarzschild Riemann
tensor, but to use the Ads-Schwarzschild metric instead of the AdS
metric.  So write
\begin {equation}
  R_{\bul\bul\bul\bul} =
  R^{(0)}_{\bul\bul\bul\bul} + (\Delta R)_{\bul\bul\bul\bul} ,
\end {equation}
where
\begin {equation}
   R^{(0)}_{IJKL}
   = \frac{1}{\Rads^2} \, (g_{IL} g_{JK} - g_{IK} g_{JL}) ,
\label {eq:R0}
\end {equation}
and where, for $z \ll \zh$,
\begin {equation}
  (\Delta R)_{\bul\bul\bul\bul} \equiv
  R_{\bul\bul\bul\bul} - R^{(0)}_{\bul\bul\bul\bul}
  \sim \frac{\Rads^2}{\zh^4}
\end {equation}
is parametrically the same size as $C$!  We never used the tracelessness
of $C$ in our previous estimates, and so our earlier power counting of
interactions with $C$'s will also work for interactions with
$\Delta R$'s.  As a result, the only thing we have to worry about
are interactions that contain powers of $R^{(0)}$.
But adding a factor of (\ref{eq:R0}) does not produce a new type of
interaction---it just costs a power of $\alpha'$.  For example,
\begin {equation}
   {\alpha'}^4 C^4 R^{(0)} \sim \frac{{\alpha'}^4}{\Rads^2} \, C^4
   \sim \lambda^{-1/2} \alpha'^3 C^4 .
\end {equation}
This is small compared to the usual $\alpha'^3 C^4$ term.
Similarly,
\begin {equation}
   {\alpha'}^4 C^3 (R^{(0)})^2 \sim \frac{{\alpha'}^4}{\Rads^4} \, C^3
   \sim \lambda^{-1} \alpha'^2 C^3
\end {equation}
is small compared to an $\alpha'^2 C^3$ term, which we saw in
section \ref{sec:C2C3} was small in turn.
All terms involving $R^{(0)}$ will be small by powers of $\lambda^{-1/2}$
compared to corrections that we have previously analyzed.


\subsection {Corrections involving the 5-form \boldmath$F$}
\label{sec:F}

The (AdS$_5$-Schwarzschild)$\times S^5$ background involves a non-vanishing
self-dual 5-form field strength $F$.
So we should check if we can get
any important corrections from terms like $D^{2n} R^m F^{2k}$
or $D^{2n+1} R^m F^{2k+1}$
\cite{Paulos}
in the supergravity Lagrangian---that is, by adding factors of
$\alpha' \bar F^2$ or $\alpha' \bar F D$
to the dispersion relations we have considered
previously.
The background value is
\begin {equation}
   \bar F_{IJKLM} = \frac{1}{\Rads} \sqrt{-g^{(5)}} \, \epsilon_{IJKLM}
\label {eq:barF}
\end {equation}
in the AdS$_5$-Schwarzschild space (and the dual of this on $S^5$),
where $\epsilon_{IJKLM}$ is the 5-component anti-symmetric
symbol.
Because of our choice back in section \ref{sec:sourceO} to choose
a source operator that is dual to components of $h_{ab}$ which
do not mix with $F$, the only relevance of $F$ in our problem is its
background value $\bar F$.  For simplicity,
we will focus on the terms where the
indices of $\bar F$ live in AdS$_5$-Schwarzschild rather than $S^5$.

An even number of factors of $\bar F$ then means
an even number of factors of 5-dimensional $\sqrt{-g} \, \epsilon$,
and an even number of such factors may always be rewritten in terms of
the metric tensor:
\begin {equation}
   -g^{(5)}
  \epsilon_{\bul\bul\bul\bul\bul}
  \epsilon_{\bul\bul\bul\bul\bul}
  \propto
  g_{\bul\bul} g_{\bul\bul} g_{\bul\bul} g_{\bul\bul} g_{\bul\bul}
  ~\mbox{(appropriately symmetrized)} .
\end {equation}
Because of this, the power counting involved
in adding a factor of $\alpha' F^2$ is similar to the previous
discussion in section \ref{sec:R0} of adding a factor of
$\alpha' R^{(0)}$, and so all such corrections are suppressed compared to
the other corrections previously considered.

What about correction terms with an odd number of $F$'s, such
as $D^{2n+1} R^m F^{2k+1}$?
The jet-stopping problem we
have chosen to study is reflection invariant in
the spatial dimensions ($x^\one, x^\two)$
transverse to the direction of the jet,
e.g.\ $x^\one \leftrightarrow x^\two$.
Specifically, the background AdS$_5$-Schwarzschild metric has this
invariance and, by choosing the source to be invariant
as in (\ref{eq:LambdaL}), the
5-dimensional response $\phi(x^\mu,x^\five)$ will be invariant as well.
So the supergravity action will only be relevant for the case where $F$
takes on its background value and all the other fields are
transverse-reflection invariant in $(x^\one,x^\two)$.  But, with these
restrictions, a term in the action with an odd number
of $\epsilon_{IJKLM}$'s from an odd number of $\bar F$'s
will contract to zero.  And so such terms may be ignored for our problem.


\end {document}